\documentclass[nature, amsfonts, amssymb, amsmath, reprint, showkeys, nofootinbib, twoside, superscriptaddress, aps, longbibliography]{revtex4-2}
\usepackage[english]{babel}
\usepackage[utf8]{inputenc}
\usepackage{amsmath}
\usepackage{siunitx}

\usepackage[colorlinks = true,
            linkcolor = blue,
            urlcolor  = blue,
            citecolor = blue,
            anchorcolor = blue]{hyperref}
\usepackage{amsthm}
\usepackage{mathtools}
\usepackage{physics}
\usepackage{xcolor}
\usepackage{graphicx}
\usepackage[left=23mm,right=13mm,top=35mm,columnsep=15pt]{geometry} 
\usepackage{adjustbox}
\usepackage{placeins}
\usepackage[T1]{fontenc}
\usepackage{lipsum}
\usepackage{csquotes}
\usepackage{xcolor}

\hypersetup{
    colorlinks=true,
    linkcolor=red,
    filecolor=magenta,      
    urlcolor=cyan,
}
\usepackage{xcolor}
\usepackage{cleveref} 

\usepackage{tikz}

\usepackage[normalem]{ulem} 

\makeatletter
\def\dketbra#1{\def\tempa{#1}\futurelet\next\dketbra@i}
\def\dketbra@i{\ifx\next\bgroup\expandafter\dketbra@ii\else\expandafter\dketbra@end\fi}
\def\dketbra@ii#1{| \tempa \rangle\!\rangle\!\langle\!\langle #1 |}
\def\dketbra@end{| \tempa \rangle\!\rangle\!\langle\!\langle \tempa |}
\makeatother

\makeatletter
\def\dbraket#1{\def\tempa{#1}\futurelet\next\dbraket@i}
\def\dbraket@i{\ifx\next\bgroup\expandafter\dbraket@ii\else\expandafter\dbraket@end\fi}
\def\dbraket@ii#1{\langle\!\langle \tempa | #1 \rangle\!\rangle}
\def\dbraket@end{\langle\!\langle \tempa | \tempa \rangle\!\rangle}
\makeatother
\usepackage[compact]{titlesec}

\makeatletter
\def\maketitle{
\@author@finish
\title@column\titleblock@produce
\suppressfloats[t]}
\makeatother

\begin{document}

\title{High-Fidelity Qutrit Entangling Gates for Superconducting Circuits}

\author{Noah Goss}
\thanks{Correspondence should be addressed to \href{mailto:noahgoss@berkeley.edu}{noahgoss@berkeley.edu}}
\affiliation{Department of Physics, University of California, Berkeley, Berkeley CA 94720, USA.}
\affiliation{Computational Research Division, Lawrence Berkeley National Laboratory, Berkeley, California 94720, USA.}
\author{Alexis Morvan}
\affiliation{Computational Research Division, Lawrence Berkeley National Laboratory, Berkeley, California 94720, USA.}
\author{Brian Marinelli}
\affiliation{Department of Physics, University of California, Berkeley, Berkeley CA 94720, USA.}
\affiliation{Computational Research Division, Lawrence Berkeley National Laboratory, Berkeley, California 94720, USA.}
\author{Bradley K. Mitchell}
\affiliation{Department of Physics, University of California, Berkeley, Berkeley CA 94720, USA.}
\author{Long B. Nguyen}
\affiliation{Computational Research Division, Lawrence Berkeley National Laboratory, Berkeley, California 94720, USA.}
\author{Ravi K. Naik}
\affiliation{Department of Physics, University of California, Berkeley, Berkeley CA 94720, USA.}
\affiliation{Computational Research Division, Lawrence Berkeley National Laboratory, Berkeley, California 94720, USA.}
\author{\\Larry Chen}
\affiliation{Department of Physics, University of California, Berkeley, Berkeley CA 94720, USA.}
\author{Christian Jünger}
\affiliation{Computational Research Division, Lawrence Berkeley National Laboratory, Berkeley, California 94720, USA.}
\author{John Mark Kreikebaum}
\affiliation{Department of Physics, University of California, Berkeley, Berkeley CA 94720, USA.}
\affiliation{Materials Science Division, Lawrence Berkeley National Laboratory, Berkeley, California 94720, USA.}
\author{David I. Santiago}
\affiliation{Computational Research Division, Lawrence Berkeley National Laboratory, Berkeley, California 94720, USA.}
\author{Joel J. Wallman}
\affiliation{Keysight Technologies Canada, Kanata, ON K2K 2W5, Canada.}
\author{Irfan Siddiqi}
\affiliation{Department of Physics, University of California, Berkeley, Berkeley CA 94720, USA.}
\affiliation{Computational Research Division, Lawrence Berkeley National Laboratory, Berkeley, California 94720, USA.}
\affiliation{Materials Science Division, Lawrence Berkeley National Laboratory, Berkeley, California 94720, USA.}
\date{\today}

\begin{abstract}
Ternary quantum information processing in superconducting devices poses a promising alternative to its more popular binary counterpart through larger, more connected computational spaces and proposed advantages in quantum simulation and error correction. Although generally operated as qubits, transmons have readily addressable higher levels, making them natural candidates for operation as quantum three-level systems (qutrits). Recent works in transmon devices have realized high fidelity single qutrit operation. Nonetheless, effectively engineering a high-fidelity two-qutrit entanglement remains a central challenge for realizing qutrit processing in a transmon device. In this work, we apply the differential AC Stark shift to implement a flexible, microwave-activated, and dynamic cross-Kerr entanglement between two fixed-frequency transmon qutrits, expanding on work performed for the $ZZ$ interaction with transmon qubits. We then use this interaction to engineer efficient, high-fidelity qutrit CZ$^\dag$ and CZ gates, with estimated process fidelities of 97.3(1)\% and 95.2(3)\% respectively, a significant step forward for operating qutrits on a multi-transmon device.
\end{abstract}
\maketitle
\section*{Introduction}
Quantum error correction (QEC) \cite{https://doi.org/10.48550/arxiv.2111.08894} is necessary for noisy intermediate-scale quantum (NISQ) \cite{preskill2018quantum} computers to realize their full potential. The surface code \cite{surface1,surface2} using qubits is considered the main route to fault tolerance \cite{ai2021exponential,marques2021logical,krinner2021realizing,zhao2022realization}, though its technical challenges have led the community to explore other approaches that could have more favorable QEC schemes, such as storing a two level system in the large Hilbert space of quantum oscillators \cite{campagne2020quantum, grimm2020stabilization}. Another alternative is to use $d$-dimensional quantum objects, or qudits, which mobilize a larger and more connected computational space than their qubit counterparts. Qutrits, the simplest form of qudits, can provide advantages in QEC for magic state distillation \cite{PhysRevX.2.041021,PhysRevLett.113.230501}, compactly encoding qubits \cite{PhysRevLett.116.150501}, and can be used to encode logical qutrits themselves \cite{PhysRevA.97.052302, Muralidharan_2017, 8248788}. Additionally, there are several proposals utilizing qutrits to improve quantum applications such as factoring with Shor's algorithm \cite{PhysRevA.96.012306}, performing the quantum Fourier transformation \cite{PhysRevA.103.032417}, providing speedups for oracle based quantum algorithms \cite{singlequdit}, improving quantum simulation \cite{sqed-simulation}, and asymptotically improving algorithms such as Grover's search \cite{10.1145/3307650.3322253, PhysRevLett.94.230502}. Realizing multi-qudit systems, however, is challenging due to the complexities of the larger Hilbert space. Nonetheless, coherent control of qudits has been performed in several physical platforms \cite{ringbauer2022,ringbauer2021universal,lanyon2008manipulating,photonic-qudit,PhysRevLett.105.223601, scrambling}.
While state of the art experiments have demonstrated high-fidelity qudit entangling gates with trapped ions \cite{ringbauer2022, ringbauer2021universal} and photonic circuits \cite{photonic-qudit}, generating high-fidelity, maximally entangling two-qudit gates remains a major challenge in superconducting circuits.

The most commonly used qubit in superconducting circuits \cite{blais2021circuit}, the transmon \cite{PhysRevA.76.042319}, is well suited to be operated as a qutrit due to its weak anharmonicity. Technical advancements in microwave engineering and improved fabrication techniques have increased transmon coherence times \cite{irfan-materials}, enabling coherent control of the full qutrit Hilbert space. Furthermore, dispersive readout can be used for high-fidelity single shot qutrit readout \cite{PhysRevLett.105.223601}. In addition, high-fidelity single qutrit operations \cite{PhysRevLett.126.210504, PhysRevLett.125.180504}, quantum information scrambling \cite{scrambling}, compact decompositions of multi-qubit gates \cite{ lanyon2009simplifying,fedorov2012implementation, song2017continuous, https://doi.org/10.48550/arxiv.2108.01652}, and improved qubit readout \cite{esp} have all been demonstrated using transmons as qutrits. Nonetheless, past qutrit entangling gates in transmons have been limited by relying on either a slow, static interaction which can only be sped up at the expense of increased quantum crosstalk on the qutrit processor or an interaction that restricts the entanglement to only a subspace of the qutrit.

In this work, we characterize the differential AC Stark shift \cite{brad, sizzle, PhysRevResearch.4.023040, PhysRevA.102.062408} on two fixed frequency transmon qutrits with static coupling and leverage it to generate dynamic qutrit entangling phases. The tunable nature of this entangling interaction enables a large on/off ratio, allowing for future high-fidelity, simultaneous single-qutrit and two-qutrit operations in transmon qutrit processors. With this interaction, we engineer the ternary controlled-Z gate (CZ) and its inverse (CZ$^\dag$). Both gates performed in our work are universal for ternary computation, maximally entangling, and Clifford gates needed for QEC in qutrits. We achieve an estimated process fidelity of 97.3(1)\% and 95.2(3)\%  for the CZ$^\dag$ and CZ respectively, measured using cycle benchmarking \cite{cb} and our generalization of the cross-entropy benchmarking routine \cite{XEB}.  The fidelity of the CZ$^\dag$ represents a factor of 4 reduction in infidelity over previous two qutrit transmon gates \cite{scrambling,PhysRevLett.126.210504}. Finally, we numerically demonstrate that our gate scheme is efficient for generating additional two-qutrit Clifford gates.
\section*{Results}
\section*{Differential AC Stark Shift}
\begin{figure}[t!]
    \centering
    \includegraphics[width = 0.9\columnwidth]{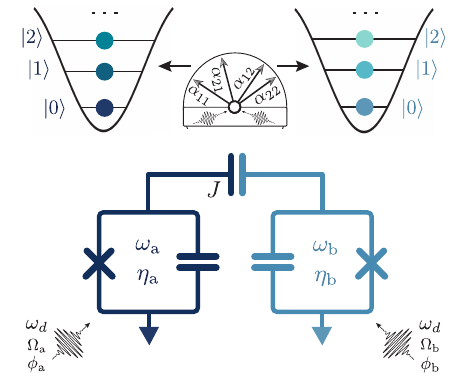}
    \caption{\textbf{Microwave-activated cross-Kerr entanglement}. Two transmon qutrits with qubit frequency $\omega_i$, anharmonicity $\eta_i$, and coupling $J$, experience a dynamical cross-Kerr ($ZZ$-like) entanglement when simultaneously driven by an off-resonant microwave drive. The strength of the cross-Kerr entangling terms ($\alpha_{11},\alpha_{12},\alpha_{21},\alpha_{22}$) is tuned by the parameters of the microwave drive ($\omega_d, \Omega, \phi)$.}
    \label{fig:1}
\end{figure}

\begin{figure*}
    \centering
    \includegraphics[width = 0.9\textwidth]{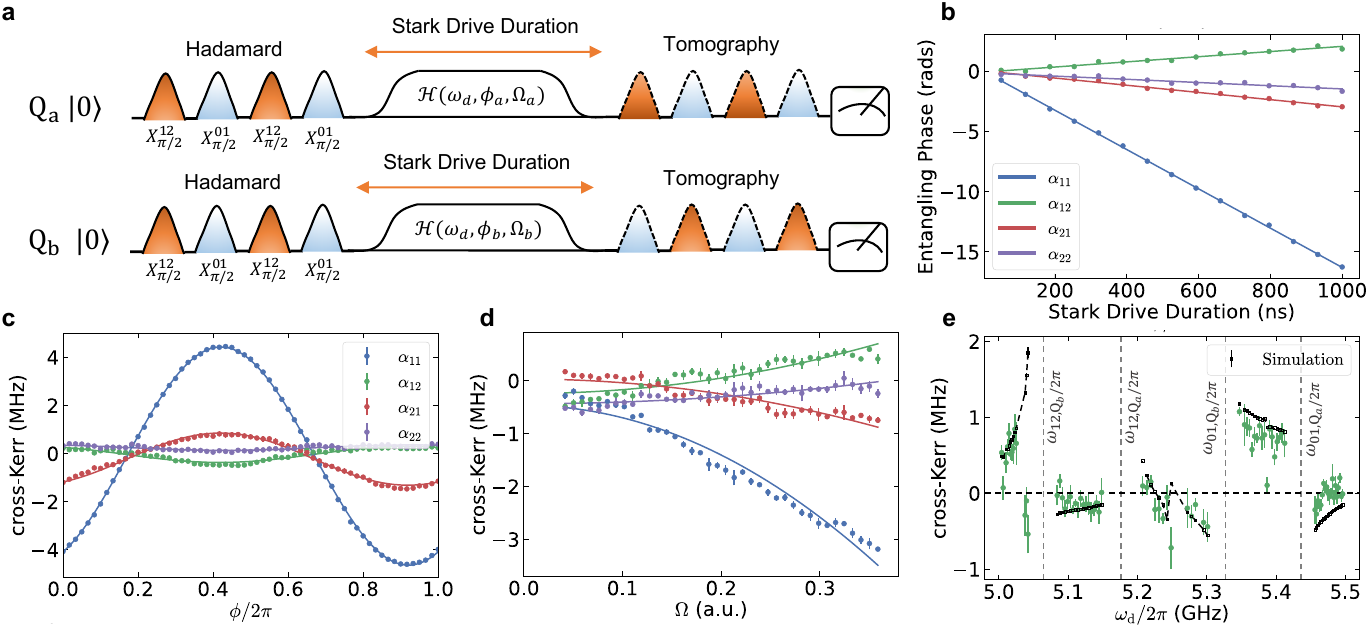}
    \caption{\textbf{Characterizing the dynamical cross-Kerr entanglement}. \textbf{a}, To study the accumulation of entangling phases under the driven cross-Kerr interaction, we place two qutrits in a full superposition using ternary Hadamard gates (virtual Z gates ommited in diagram), then study the evolution under the Stark drive scheme by performing state tomography. \textbf{b}, We demonstrate fitting the accumulation of entangling phase found by tomography to our linear, driven cross-Kerr model, where $\alpha_{ij}$ is the slope of the line and the uncertainty is from the linear fit. \textbf{c-d}, We match the behavior of the cross-Kerr entanglement given relevant experimental parameters in our system to our Hamiltonian model for the relative phase of the driving, $\phi$, and amplitude of the driving, fixing $\Omega = \Omega_a = \Omega_b$. \textbf{e}, We additionally compare the dependence of $\alpha_{12}$ on the frequency of the drive $\omega_d$ using an \textit{ab-initio} master equation simulation in QuTiP \cite{JOHANSSON20121760,JOHANSSON20131234}.}
    \label{fig:tunable_kerr}
\end{figure*}

Recent works by Refs.~\cite{brad, sizzle, PhysRevResearch.4.023040, PhysRevA.102.062408} demonstrated that the architecture employed in the Cross-resonance (CR) entangling gate can also realize a two-qubit CZ gate by leveraging the conditional Stark shifts from simultaneously driving a pair of coupled qubits off-resonantly. The advantages of this method are two-fold: firstly, the CZ gate commutes with $ZZ$ errors from the always-on dispersive coupling between the transmons. Secondly, unlike in the CR gate, the frequency of the microwave drive can take a range of values. This flexibility in drive frequency affords significant advantages in avoiding frequency collisions with other transmons or spurious two-levels systems \cite{PhysRevResearch.4.023079}. The generalization of conditional Stark shifts to qutrits is straightforward. Working in the energy eigenbasis of our two qutrit Hilbert space, up to single-qutrit phases, one's system evolves under the cross-Kerr Hamiltonian:
\begin{equation}
    \label{eq:H_ck}
    \begin{gathered}
    \mathcal{H} = \alpha_{11}\ket{11}\bra{11} +  \alpha_{12}\ket{12}\bra{12}  \\ +  \alpha_{21}\ket{21}\bra{21} +  \alpha_{22}\ket{22}\bra{22},
    \end{gathered}
\end{equation}
where each term can be be calculated with perturbation theory (see supplementary note 2). In this microwave-activated case, the $\alpha_{ij}$ ($ZZ$-like) terms are given by:
\begin{equation}
    \alpha_{ij} = A_{ij}(\omega_d) \Omega_a \Omega_b \cos(\phi_a - \phi_b),
    \label{eq:cross-kerr}
\end{equation}
where $\Omega_i$ and $\phi_i$ are respectively the amplitude and phsae of the drive on transmon $i$. The coefficients $A_{ij}$ are functions of the proximity of the microwave drive frequency ($\omega_d$) to nearby transitions. We note that this Hamiltonian generates entanglement between the entire two-qutrit Hilbert space, contrary to the CR case, where the entanglement is mostly restrained to a subspace of the qutrit \cite{scrambling, PhysRevApplied.17.024062}. This dynamic, driven cross-Kerr interaction is depicted schematically in Fig.~\ref{fig:1}. It is important to note that the number of degrees of freedom in this interaction are not sufficient, in general, to realize a Clifford two-qutrit gate like the CZ gate with a single round of cross-Kerr entanglement, a difficulty discussed in further detail in the next section.

Measuring the $ZZ$ interaction in the qubit case can be performed by a conditional Ramsey experiment or through a dynamically decoupled JAZZ (Joint-Amplification-of-$ZZ$) sequence that removes the low frequency drift \cite{PhysRevLett.119.180501,648}. In the larger Hilbert space of two qutrits, we need to measure four of these entangling phases with a rate of accumulation set by $\alpha_{ij}$ in Eq.~\ref{eq:cross-kerr}. To simplify the measurement and reduce the number of experiments needed, we generalize the controlled-Ramsey experiment to the full qutrit space with a pulse sequence presented in Fig.\,\ref{fig:tunable_kerr}a. In this sequence, we apply simultaneous ternary Hadamard gates on both qutrits, execute the microwave drive, and subsequently perform the full two qutrit state tomography. Doing so for several durations of the Stark driving allow us to fully characterize the conditional and unconditional Stark shifts.

We demonstrate in Fig.\,\ref{fig:tunable_kerr}b the result of this measurement scheme: the entangling phases increase linearly with the duration of the Stark drive, where the proportionality constant is set by $\alpha_{ij}$ as predicted by our cross-Kerr model in Eq.~\ref{eq:H_ck}. In Fig.\,\ref{fig:tunable_kerr}c-d, we present how the driven cross-Kerr interaction depends on the parameters of our entanglement scheme, specifically the phase and the amplitude of the Stark drive. We note that the qualitative behavior is properly captured by the perturbation theory in Eq.~\ref{eq:cross-kerr}. We also explore the behavior as a function of the drive frequency in Fig.\,\ref{fig:tunable_kerr}e. In this case, the perturbation theory fails, but an \textit{ab-initio} master equation simulation captures some of the response; additional details on the frequency dependence of all $\alpha_{ij}$ terms and the master equation simulation can be found in the supplement. We expect the unaccounted features can be attributed to higher-order terms, frequency dependent classical crosstalk, or parasitic two level systems (TLS) in our device. In an experimental setting, the flexibility of this entanglement allows us the freedom to set the drive frequency far from any of these features.

\section*{Qutrit CZ/CZ$^\dag$ gate}

\begin{figure}[t!]
    \centering
    \includegraphics[width = \columnwidth]{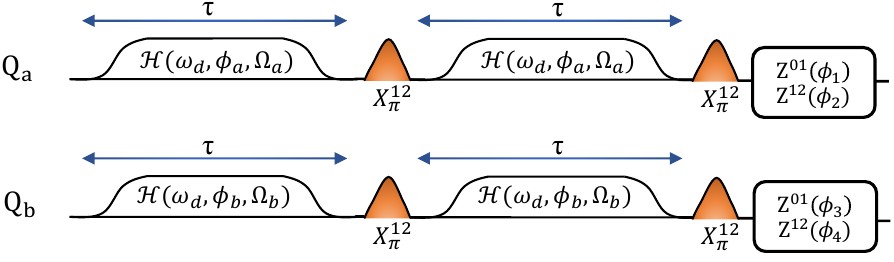}
    \caption{\textbf{Gate schematic}. For the CZ and CZ$^\dag$ gate, we perform two rounds of cross-Kerr entanglement for duration $\tau$ with interleaved echo pulses in the $\{ \ket{1},\ket{2}\}$ subspace which shuffle the entangling phases. For proper conditions on the $\alpha_{ij}$ terms in Eq.~\ref{eq:H_ck}, the CZ$^\dag$(CZ) is compiled with a total gate time of 580(783) ns. The local Z terms in both two level subspaces of the qutrit are then undone using virtual Z gates.}
    \label{fig:gate_schematic}
\end{figure}

We next construct qutrit controlled-phase gates utilizing this entangling interaction. The qutrit CZ and CZ$^\dag$ gate are both maximally entangling and members of the two-qutrit Clifford group making them particularly useful gates for ternary computation. The CZ gate is defined as:
\begin{equation}
    U_{\textnormal{CZ}} = \sum_{i, j \in \{0,1,2\}^2} \omega^{ij} \ketbra{ij}{ij},
    \label{eq: CZD}
\end{equation}
with $\omega = e^{2i\pi/3}$, the third root of unity; the CZ unitary follows directly from generalizing the qubit Pauli group to qutrits and is explained in further detail in Supplementary Note 3. Under simultaneous Stark drives, the two-qutrit Hilbert space follows the unitary evolution $
U = \textnormal{exp}(-i(\mathcal{H} + \phi_1 I\otimes Z^{01} + \phi_2 I\otimes Z^{12} + \phi_3 Z^{01}\otimes I + \phi_4 Z^{12} \otimes I)\tau)$ where $\mathcal{H}$ is given in Eq.~\ref{eq:H_ck}. To perform a CZ gate with a single round of cross-Kerr driving, for example, one would need to find driving parameters meeting the conditions: $\{ \alpha_{11} = \alpha_{22} = 2\alpha_{21} = 2\alpha_{12} \}$, a task that is not broadly feasible. Practically speaking, we desire a compromise between the most general and robust gate scheme and this ``fine-tuned” approach, while still taking advantage of the dynamical nature of our cross-Kerr interaction. By employing the pulse scheme in Fig.~\ref{fig:gate_schematic}, where echo pulses in the $\{\ket{1},\ket{2}\}$ subspace shuffle entangling phases, we have the modified unitary evolution (omitting the single-qutrit phases for brevity):
\begin{equation}
\begin{gathered}
    U =  \textnormal{exp}(-i[(\alpha_{11} + \alpha_{22})\tau(\ketbra{11}{11}  + \ketbra{22}{22} )+ \\ (\alpha_{12} + \alpha_{21})\tau(\ketbra{12}{21} + \ketbra{21}{21}) ]),
\end{gathered}
\label{evol}
\end{equation} 

Using the experimental knobs demonstrated in Fig.~\ref{fig:tunable_kerr}, we are able to satisfy approximate conditions on our cross-Kerr evolution that are ideal for compactly generating qutrit controlled phase gates. Specifically, we find drive parameters that satisfy $(\alpha_{11} + \alpha_{22}) \approx -(\alpha_{12} + \alpha_{21})$ for performing the CZ gate and $(\alpha_{11} + \alpha_{22}) \approx 2(\alpha_{12} + \alpha_{21})$ for performing the CZ$^\dag$ gate.The Stark drive parameters that meet these conditions are provided in Supplementary Note 1. Under these conditions on the cross-Kerr, and with the unitary evolution provided by our gate scheme in Eq. \ref{evol}, at some drive time $\tau$, we will have approximately acquired the desired entangling phases found in Eq.\ \ref{eq: CZD} to synthesize respectively a CZ or CZ$^\dag$ unitary. To ensure adiabaticity and limit leakage, we perform the Stark drive via flat-top cosine pulses with ramp up and down features. This ramp leads to effective offsets on the accumulation of the entangling phases, as we only expect the linear accumulation of entangling phase given by $\alpha_{ij}$ in Eq.\ \ref{eq:cross-kerr} to correspond to the flat-top of the Stark drive. When tuning up the two-qutrit gates, this means that we first perform parameters sweeps to find regions where the previously mentioned conditions on the $\alpha_{ij}$s are met, then perform the actual pulse scheme in Fig. \ref{fig:gate_schematic} and adjust the Stark drive parameters until the target entangling phases given by Eq.\ \ref{eq: CZD} are most accurately acquired. Finally, as outlined schematically in Fig.~\ref{fig:gate_schematic}, one can undo the local Z phases (found via tomography) in both the \{$\ket{0}$,$\ket{1}$\} and \{$\ket{1}$,$\ket{2}$\} subspaces with virtual Z gates \cite{PhysRevA.96.022330}. In this work, we performed the CZ and CZ$^\dag$ on two different pairs of transmon qutrits, demonstrating the flexible nature of generating two-qutrit gates from this driven cross-Kerr scheme.

\section*{Benchmarking \label{sec:benchmarking}}

\begin{figure}[h!]
    \centering
    \includegraphics[width = \columnwidth]{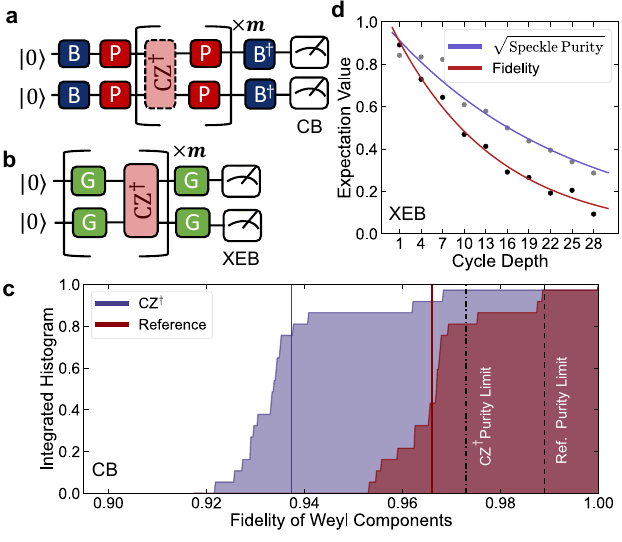}
    \caption{\textbf{Benchmarking}. \textbf{a}, Circuit schematic of cycle benchmarking (CB). The errors of the CZ$^\dag$ are twirled via random Weyl gates (red) to tailor errors into stochastic Weyl channels. The initial state and measurement basis (blue) are selected to pick out the decay associated with specific Weyl channels. \textbf{b}, Circuit schematic of cross-entropy benchmarking (XEB). The errors of the CZ$^\dag$ are twirled via random SU(3) gates (green) to tailor the noise to a simple depolarizing channel. \textbf{c}, An integrated histogram of CB for both the CZ$^\dag$ gate and a reference cycle, with the solid vertical lines giving the fidelities 0.936(1) and 0.966(1) respectively, yielding an estimated process fidelity of 97.3(1)\%. We extract an error budget directly from CB, estimating a purity limited fidelity of 0.973(9) and 0.989 (with negligible error) for the dressed CZ$^\dag$ and reference cycles, yielding a purity limit 0.986(9) for the isolated CZ$^\dag$ gate. \textbf{d}, From XEB we estimate the depolarized fidelity as 0.933(3). Additionally, we estimate the speckle-purity limited fidelity of the CZ$^\dag$ dressed with random SU(3) gates to be 0.961(3). }
    \label{fig:gate_benchmarking}
\end{figure}
We first benchmark our two-qutrit gates with cycle benchmarking (CB)~\cite{cb} using True-Q~\cite{true-q_2020}. While originally written in terms of qubits, CB naturally generalizes to qutrits~\cite{PhysRevLett.126.210504}. We use CB instead of, e.g., interleaved randomized benchmarking~\cite{PhysRevLett.106.180504,PhysRevLett.109.080505}, because it requires significantly fewer multi-qutrit gates per circuit. We describe the generalization in the supplementary material. With this technique, we estimate the Weyl (generalized-Pauli) error rate of the CZ$^\dag$ and CZ gate to be $2.7(1)\%$ and $4.8(3)\%$ respectively. By contrast, the highest fidelity, two-qutrit gate performed previously with transmons had an error rate of 11.1\% \cite{scrambling}. CB also allows us to construct the Weyl-twirled error per channel \cite{PhysRevLett.126.210504} of the unitary in Fig.\,\ref{fig:gate_benchmarking}. This provides us with an estimate of the worst case scenario of less than $8\%$ and demonstrates a relatively low dispersion of our error channels.

As an added confirmation of the fidelity of the CZ$^\dag$ gate, we generalize the cross-entropy benchmarking (XEB) routine \cite{XEB, MullaneXEBTheory} to tailor all gate errors into a depolarizing channel, for work with qutrit unitaries. In the qutrit case, we find sufficient tailoring of our noise can be performed by interleaving random SU(3) gates around our target gate. The circuit diagram for qutrit XEB is in Fig.\,\ref{fig:gate_benchmarking}b and the results can be found in Fig.\,\ref{fig:gate_benchmarking}e. We find that the depolarized fidelity of the CZ$^\dag$ dressed with random SU(3) gates agrees with the estimate of the process fidelity from our Weyl twirled CB results within a standard error. Additional discussion of the qutrit XEB method is provided in the supplementary materials. 

Finally, we would like to be able to characterize what fraction of the errors present in our two qutrit unitaries are coherent on the time scale of multiple experiments, and thus could be removed by improved calibration. 
As we show in supplementary note 6, under the depolarizing unitary noise model, the variance of CB and XEB circuits both provide a robust method of estimating the purity limit~\cite{Wallman2015,arute2019quantum}.
The corresponding estimates are shown in  Fig.\ref{fig:gate_benchmarking}c with the CB estimate of $97.3(9)\%$ exceeding the speckle-purity limit of $96.1(3)\%$ for the dressed CZ$^\dag$ gate. This disagreement is likely due to the fact that the CB data reveals that the noise is dominated by single-qutrit phase errors. As these errors are likely to fluctuate around a mean, they will add dephasing errors that are not captured by the depolarizing unitary model used to analyze XEB. Another possible contributing factor is that the noise fluctuated between the XEB and CB experiments, which were performed in separate batches.

\section*{Gate synthesis of two qutrit unitaries}
\begin{figure}[]
    \centering
    \includegraphics[width = \columnwidth]{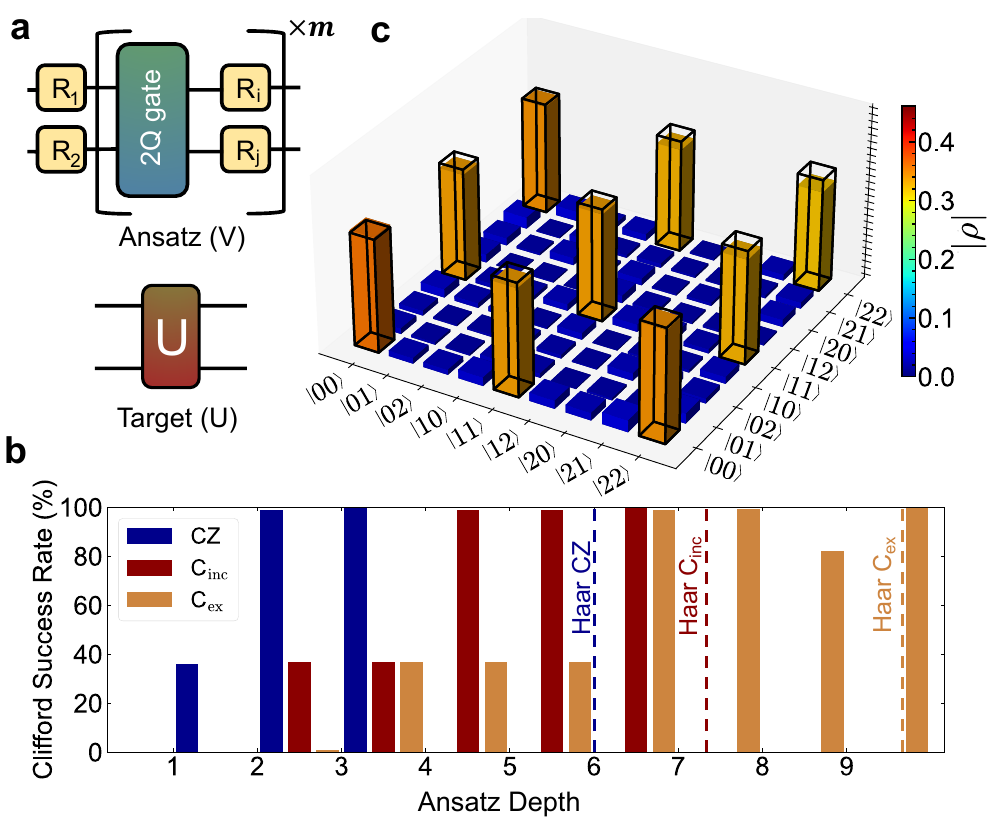}
    \caption{\textbf{Demonstration of gate expressability}. \textbf{a}, A parameterized Ansatz circuit (V) is used to synthesize a target unitary (U), given some 2-qutrit gate and arbitrary SU(3) gates. \textbf{b}, We study the Ansatz circuit (V) for the different two-qutrit gates discussed in the text for 1000 Haar random and Clifford gates, minimizing the infidelity $1-\mathcal{F}$(V,U). The dashed lines represent 100\% numerical success for synthesizing our set of Haar random gates, and the bars display the success rate for synthesizing Clifford gates. We perform the minimization until we find a 100\% success rate for each two-qutrit gate between depths $0 \leq m \leq 9$. \textbf{c}, An experimentally reconstructed density matrix of the two qutrit Bell state $\ket{\psi} = \frac{1}{\sqrt{3}}(\ket{00} + \ket{11} + \ket{22})$ formed using a single CZ gate with state fidelity $\mathcal{F}=0.952$. The black outline is the target density matrix.}
    \label{fig:gate_pow}
\end{figure}
To study the expressibility of the two-qutrit gates in this work, we numerically explore the ability of the ternary CZ/CZ$^\dag$ (localy equivalent to each other and the CSUM gate) to synthesize other two-qutrit gates, and compare them to two-qutrit entangling gates that only entangle a subspace of the qutrit, such as the controlled-exchange (C$_{\rm ex}$) and controlled-increment (C$_{\rm inc}$) gates performed on a trapped ion system in Ref.~\cite{ringbauer2021universal}. To this end, we consider an Ansatz circuit V as in Fig.\,\ref{fig:gate_pow}a, with depth $m$, which we use to synthesize target circuits belonging to either the two qutrit Clifford group or set of Haar random gates. The gate synthesis is performed by optimizing the ansatz parameters to minimize the distance between V and U, i.e. the infidelity 1-$\mathcal{F}\rm{(V,U)}$. 

We perform this numerical investigation on 1000 Haar random gates and 1000 Clifford gates. We find that all 1000 Haar random gates can be synthesized at depth 6 for CZ/CZ$^\dag$, 7 for C$_{\rm inc}$, and 9 for C$_{\rm ex}$. The synthesis success rate for all 3 unitaries in terms of target Clifford circuits are shown in Fig.\,\ref{fig:gate_benchmarking}b. Notably, almost all two qutrit Clifford gates were successfully compiled at depth 2 in CZ/CZ$^\dag$, with 100\% success at depth 3. By contrast, C$_{\rm ex}$ and C$_{\rm inc}$ did not demonstrate as much improvement at synthesizing Clifford gates over Haar random gates, achieving 100\% success for target Clifford gates at depth 6 for C$_{\rm inc}$ and 9 for C$_{\rm ex}$. Additionally, unique amongst these gates, the CZ/CZ$^\dag$ can generate maximally entangled two qutrit states with a single iteration of the gate. We demonstrate the power of this feature in the experimentally reconstructed qutrit Bell-state density matrix in Fig.\,\ref{fig:gate_pow}c. In summary, the maximally entangling CZ/CZ$^\dag$ gates have low intrinsic errors and can also can synthesize a very important family of gates for QEC (the Cliffords) \cite{https://doi.org/10.48550/arxiv.2111.08894} at much lower depths than the two-qutrit gates which only entangle a subspace of the qutrit.

\section*{Discussion}
We realized a microwave-activated, dynamic cross-Kerr entangling interaction that can be employed to engineer qutrit entangling phases with high precision. Leveraging this interaction, we generated two maximally entangling and high-fidelity two-qutrit gates on two separate pairs of fixed-frequency transmon qutrits. We demonstrated numerically that these two qutrit gates are efficient for producing additional two-qutrit unitaries, especially other Clifford gates. Future work developing a systematic gate tune up procedure may prove essential in improving the fidelity and scalability of our approach. Additionally, a study of the effects of this gate scheme on spectator qutrits will also be necessary for determining its scalability. We expect that by being maximally entangling and a member of the two-qutrit Clifford group, the gates performed in this work will prove especially powerful in future efforts to employ qutrits for QEC, quantum simulation, and quantum computation. Perhaps most importantly, all of this work was performed on multi-transmon quantum processors which are normally used for qubit experiments; the untapped potential of transmons as qutrits is only beginning to be explored. As a final note, the two-qutrit Hilbert space is larger than even the three qubit Hilbert space; as the community continues to explore qudits, we propose that metrics and benchmarks should be developed to reasonably compare qudit vs. qubit gates.

\section*{Data Availability}
The data that support the ﬁndings of this study are available from the corresponding authors on reasonable request.

\section*{Code Availability}
Cycle Benchmarking and the expressability analysis were performed using properietary TrueQ\textsuperscript{TM} software (https://trueq.quantumbenchmark.com). All other code that supports the ﬁndings of this study is available from the corresponding authors on reasonable request.


\bibliography{ms}
\newpage
\section*{Acknowledgements}
We are grateful to A. Hashim, W. Livingston, and I. Hincks for conversations and insights. This work was supported by the Quantum Testbed Program of the Advanced Scientific Computing Research Division, Office of Science of the U.S. Department of Energy under Contract No. DE-AC02-05CH11231.

\section*{Author Contributions}
N.G. and A.M. conceived and planned the experiment. N.G., A.M., B.M., and B.K.M. performed the experiment. N.G., B.M., and J.J.W. analysed the data. L.B.N. and R.K.N. contributed to the analysis and discussion of the results. L.C., C.J., and J.M.K. fabricated the transmon devices. N.G. wrote the manuscript with assistance from A.M., J.J.W., B.M., L.B.N., and R.K.N. All work was carried out under the supervision of D.I.S. and I.S.

\section*{Competing Interests}
J.J.W. has a financial interest in Keysight technologies and the TrueQ\textsuperscript{TM} software. All other authors declare no competing interests.
\clearpage
\onecolumngrid



\end{document}


\title{Supplementary Information: High-Fidelity Qutrit Entangling Gates for Superconducting Circuits}

\author{Noah Goss}
\affiliation{Department of Physics, University of California, Berkeley, Berkeley CA 94720, USA.}
\affiliation{Computational Research Division, Lawrence Berkeley National Laboratory, Berkeley, California 94720, USA.}
\author{Alexis Morvan}
\affiliation{Computational Research Division, Lawrence Berkeley National Laboratory, Berkeley, California 94720, USA.}
\author{Brian Marinelli}
\affiliation{Department of Physics, University of California, Berkeley, Berkeley CA 94720, USA.}
\affiliation{Computational Research Division, Lawrence Berkeley National Laboratory, Berkeley, California 94720, USA.}
\author{Bradley K. Mitchell}
\affiliation{Department of Physics, University of California, Berkeley, Berkeley CA 94720, USA.}
\affiliation{Computational Research Division, Lawrence Berkeley National Laboratory, Berkeley, California 94720, USA.}
\author{Long B. Nguyen}
\affiliation{Computational Research Division, Lawrence Berkeley National Laboratory, Berkeley, California 94720, USA.}
\author{Ravi K. Naik}
\affiliation{Department of Physics, University of California, Berkeley, Berkeley CA 94720, USA.}
\affiliation{Computational Research Division, Lawrence Berkeley National Laboratory, Berkeley, California 94720, USA.}
\author{\\ Larry Chen}
\affiliation{Department of Physics, University of California, Berkeley, Berkeley CA 94720, USA.}
\author{Christian Jünger}
\affiliation{Computational Research Division, Lawrence Berkeley National Laboratory, Berkeley, California 94720, USA.}
\author{John Mark Kreikebaum}
\affiliation{Department of Physics, University of California, Berkeley, Berkeley CA 94720, USA.}
\affiliation{Materials Science Division, Lawrence Berkeley National Laboratory, Berkeley, California 94720, USA.}
\author{David I. Santiago}
\affiliation{Computational Research Division, Lawrence Berkeley National Laboratory, Berkeley, California 94720, USA.}
\author{Joel J. Wallman}
\affiliation{Keysight Technologies Canada, Kanata, ON K2K 2W5, Canada.}
\author{Irfan Siddiqi}
\affiliation{Department of Physics, University of California, Berkeley, Berkeley CA 94720, USA.}
\affiliation{Computational Research Division, Lawrence Berkeley National Laboratory, Berkeley, California 94720, USA.}
\affiliation{Materials Science Division, Lawrence Berkeley National Laboratory, Berkeley, California 94720, USA.}
\date{\today}

\maketitle

\onecolumngrid
\renewcommand{\figurename}{Supplementary Fig.}
\renewcommand{\tablename}{Supplementary Table}
\setcounter{figure}{0}  
\renewcommand*\thetable{\arabic{table}}

\renewcommand{\theequation}{S\arabic{equation}}
\setcounter{equation}{0}   
\renewcommand{\thesection}{Supplementary Note \arabic{section} }

\section{Device Parameters}
Both gates and cross-Kerr characterization data was taken using an 8 transmon ring with fixed-frequency transmons and fixed-frequency coupling mediated by a coplanar waveguide resonator. The CZ and CZ$^\dag$ gate are performed on two different chips, using two different pairs of transmon qutrits. We give the relevant single-qutrit parameters and coherences for the CZ and CZ$^\dag$ gate in Table \ref{tab:single_qutrit_parameters_cz} and Table \ref{tab:single_qutrit_parameters_czd} respectively. Here, $T_{2r}$ denotes coherence statistics taken using a Ramsey experiment and $T_{2e}$ using an echo pulse. We additionally provide the approximate gate parameters for the two gates performed in the work, where $\omega_d$ denotes the drive frequency, $\Omega$ denotes the approximate drive strength, and $\tau_g$ denotes the total gate time including the $\pi$ pulses in the $\{\ket{1} \ket{2} \}$ subspace.

\begin{table}[h!]
  \centering
  \resizebox{0.3\columnwidth}{!}{
  \begin{tabular}{|l||r|r|r|r|r|r|r|r|r|r|r|}
    \hline
    {} & Q3 & Q4 \\
    \hline
    \hline
    Qubit freq.~(GHz) & 5.436 & 5.327 \\
    Anharm. (MHz) & -260.20 & -262.94 \\
    $T_1^{01}$ ($\mu$s) & 125(37) & 78(16) \\
    $T_1^{12}$ ($\mu$s) & 63(9) & 47(5) \\
    $T_{2e }^{01}$ ($\mu$s) & 190(28) & 138(25) \\
    $T_{2e}^{12}$ ($\mu$s) & 61(13) & 45(7) \\
    $T_{2e}^{02}$ ($\mu$s) & 75(19) & 62(6) \\
    $T_{2r }^{01}$ ($\mu$s) & 114(47) & 99(24) \\
    $T_{2r}^{12}$ ($\mu$s) & 17(8) & 17(9) \\
    $T_{2r}^{02}$ ($\mu$s) & 20(16) & 21(9) \\
    $\omega_d$ (GHz) & 5.287 & 5.287 \\
    $\Omega$ (MHz) & $\approx 13$ & $\approx 13$ \\
    $\tau_g$ (ns) & 783 &  \\

    \hline
  \end{tabular}}
\caption{Single-qutrit parameters for the pair of transmons used to perform the CZ gate.}\label{tab:single_qutrit_parameters_cz}
\end{table}
\begin{table}[h!]
  \centering
  \resizebox{0.3\columnwidth}{!}{
  \begin{tabular}{|l||r|r|r|r|r|r|r|r|r|r|r|}
    \hline
    {} & Q5 & Q6 \\
    \hline
    \hline
    Qubit freq.~(GHz) & 5.362 & 5.523 \\
    Anharm. (MHz) & -275 & -271.35 \\
    $T_1^{01}$ ($\mu$s) & 45(7) & 58(7) \\
    $T_1^{12}$ ($\mu$s) & 33(3) & 28(3) \\
    $T_{2e}^{01}$ ($\mu$s) & 63(7) & 84(6) \\
    $T_{2e}^{12}$ ($\mu$s) & 28(3) & 30(3) \\
    $T_{2e}^{02}$ ($\mu$s) & 37(3) & 35(3) \\
    $T_{2r }^{01}$ ($\mu$s) & 36(9) & 76(8) \\
    $T_{2r}^{12}$ ($\mu$s) & 10(6) & 18(6) \\
    $T_{2r}^{02}$ ($\mu$s) & 11(6) & 21(8) \\
    $\omega_d$ (GHz) & 5.191 & 5.191 \\
    $\Omega$ (MHz) & $\approx 11$ & $\approx 11$ \\
    $\tau_g$ (ns) & 580 &  \\

    \hline
  \end{tabular}}
\caption{Single-qutrit parameters for the pair of transmons used to perform the CZ$^\dag$ gate.}\label{tab:single_qutrit_parameters_czd}
\end{table}
\newpage

\section{Perturbation Theory}
\begin{figure}[h!]
    \centering
    \includegraphics[width = \textwidth ]{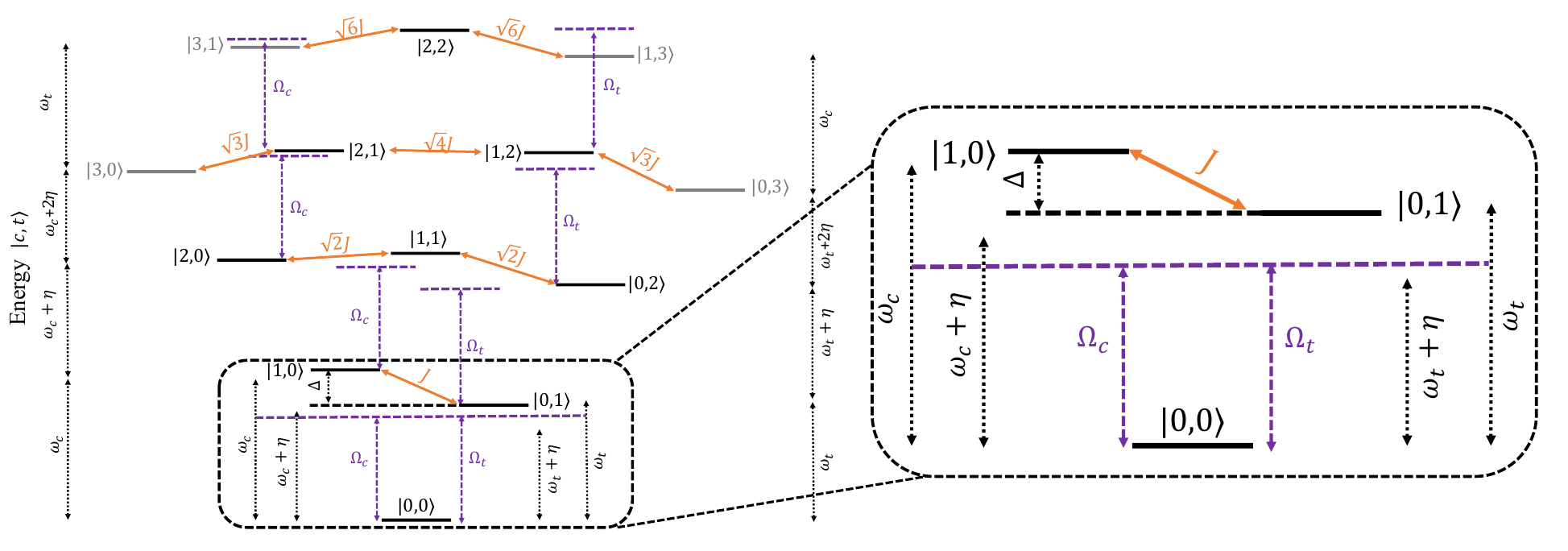}
    \caption{\textbf{Drive scheme for conditional stark-induced cross-Kerr Hamiltonian }. We present an example energy level diagram in the rotating frame of the drive where we place the off resonant microwave drive on both qutrits between their two respective $\ket{1} \rightarrow \ket{2}$ transitions. The simultaneous off resonant microwave drives induce conditional Stark shifts that generate entangling phases on four states in our two qutrit Hilbert space.}
    \label{levels_diagram}
\end{figure}
In the frame of the drive at frequency $\omega_{\text{d}}$ and after making a rotating wave approximation (RWA), the system Hamiltonian is
\begin{equation}
    \label{eq:Hsupp}
    \begin{gathered}
        H=\sum_{i=\mathrm{c,t}}\left[\left(\omega_{i}-\omega_{\mathrm{d}}\right)a_{i}^{\dag}a_{i}+\frac{\eta_{i}}{2}a_{i}^{\dag}a_{i}^{\dag}a_{i}a_{i} + \Omega_{i}\left(e^{i\varphi_{i}}a_{i}+e^{-i\varphi_{i}}a_{i}^{\dag}\right)\right]+J\left(a_{\mathrm{c}}^{\dag}a_{\mathrm{t}}+a_{\mathrm{c}}a_{\mathrm{t}}^{\dag}\right)
    \end{gathered}
\end{equation}
with $\hbar=1$ and the transmons approximated as Duffing oscillators with qubit frequency $\omega_i$, anharmonicity $\eta_i$, capacitive coupling $J$, and where we define $a_i$ as the bosonic annihilation operator. The parameters of the drive are given as amplitude $\Omega_i$, and phase $\varphi_i$. Since only the relative drive phase is physical, we choose a basis where $\varphi_{\text{c}}=0$ and $\varphi_{d}=\varphi_{t}-\varphi_{c}$. The detuning of transmon $i$ from the drive is $\Delta_{i}=\omega_{i}-\omega_{d}$. We analyze the system perturbatively in the limit $\Omega_{i}, J \ll |\eta_{i}|, |\Delta_{i}|$. In this limit the bare transmon Hamiltonians serve as the unperturbed system, $H_{0}$, and the perturbation, $V$, is composed of the single qubit drive terms and the coupling term. Time independent perturbation theory applied to Eq.~\ref{eq:Hsupp} will yield energies $E_{ij}$, the approximate diagonal elements of the Hamiltonian in the basis labelled by the transmon occupation numbers $\ket{ij}$ with $i,j=0,1,2,\dots,d-1$ the state of the ``control'' (c) and ``target'' (t) transmons respectively. The dimension where we truncate the Hamiltonian is $d$. In the present case of qutrits $d=3$ and $H' \approx \sum_{i,j=0}^{2}
\tilde{E}_{ij}\ket{ij}\bra{ij}$ where we define $\tilde{E}_{ij}=E_{ij}-E_{00}$, performing a global shift of the energies to set the energy of the $\ket{00}$ state to zero. 

We isolate the entangling cross-Kerr terms by transforming $H'$ according to the unitary 
\begin{equation}
    \begin{gathered}
    U=\text{exp} \Big\{-it\left[\tilde{E}_{01}I\otimes\ket{1}\bra{1}+\tilde{E}_{10}\ket{1}\bra{1}\otimes I +  \tilde{E}_{02}I\otimes\ket{2}\bra{2}+\tilde{E}_{20}\ket{2}\bra{2}\otimes I\right]\Big\},    
    \end{gathered}
\end{equation}
where $I=\ket{0}\bra{0}+\ket{1}\bra{1}+\ket{2}\bra{2}$ is the single qutrit identity operator. The transformation eliminates the single qutrit energies $\tilde{E}_{i0}$ and $\tilde{E}_{0j}$, $i,j=1,2$ which simply result in local phases that can be eliminated by virtual single qutrit phase gates. The transformed Hamiltonian 
\begin{equation}
    \label{eqn:HcrossKerr}
    H''=\alpha_{11}\ket{11}\bra{11}+\alpha_{21}\ket{21}\bra{21}+\alpha_{12}\ket{12}\bra{12}+\alpha_{22}\ket{22}\bra{22}
\end{equation}
is written in terms of the cross-Kerr rates $\alpha_{ij}$ which describe the rates at which the entangling phases on the states $\ket{ij}$ are accumulated. These $\alpha_{ij}$, explicitly given by
\begin{equation}
    \alpha_{ij}=E_{ij}+E_{00}-E_{0j}-E_{i0},
\end{equation}
generalize the qubit ZZ rate $\zeta$ to the case of qutrits \cite{brad}. 

In perturbation theory we expand $\alpha_{ij}=\alpha_{ij}^{(0)} + \alpha_{ij}^{(1)}+\cdots$ where $\alpha_{ij}^{(n)}=E_{ij}^{(n)}+E_{00}^{(n)}-(E_{i0}^{(n)}+E_{0j}^{(n)})$ and $E_{ij}^{(n)}$ is the $n$th order correction to the energy of state $\ket{ij}$. The first two terms $\alpha_{ij}^{(0)}$ and $\alpha_{ij}^{(1)}$ are zero because in the unperturbed system, $H_{0}$, the transmons are not interacting and the perturbation $V$ only maps states $\ket{ij}$ to states $\ket{i'j'}$ that are orthogonal to $\ket{ij}$. The first contribution comes at second order where the static coupling of strength $J$ between the transmons generates a static cross-Kerr interaction with rates
\begin{align}
    \alpha_{11}^{(2)}&=\frac{4\eta J^{2}}{(\eta-\Delta)(\eta+\Delta)} \\
    \alpha_{21}^{(2)}&=\frac{2\eta J^{2}(5\eta-4\Delta)}{\Delta(\eta-\Delta)(2\eta-\Delta)} \\
    \alpha_{12}^{(2)}&=-\frac{2\eta J^{2}(5\eta+4\Delta)}{\Delta(\eta+\Delta)(2\eta+\Delta)} \\
    \alpha_{22}^{(2)}&=\frac{16\eta J^{2}}{(\eta+\Delta)(\eta-\Delta)}
\end{align}
where $\Delta = \omega_{\text{c}}-\omega_{\text{t}}$ and we have set $\eta_{\text{c}}=\eta_{\text{t}}=\eta$ in order to arrive at more compact expressions. In most systems this is a reasonable approximation and in particular the pair of transmons used in this work have anharmonicities $\eta_{\text{c}}=272~\text{MHz}$ and $\eta_{\text{t}}=270~\text{MHz}$.

As in the qubit case the driven cross-Kerr interaction contributes starting at third order when both transmons are driven (see the discussion in the Supplementary Materials of \cite{brad}) and we find rates
\begin{equation}
    \alpha_{11} =\frac{8\cos\phi\eta^{2}\Omega_{C}\Omega_{T}J}{\Delta_{C}\Delta_{T}(\eta-\Delta_{C})(\eta-\Delta_{T})}
\end{equation}
\begin{equation}
    \alpha_{12}/\alpha_{11}= \frac{\Delta_{C}^{2}(\eta-2\Delta_{T})+\eta(2\eta^{2}-2\eta\Delta_{T}-\Delta_{T}^{2}))+\Delta_{C}(-3\eta^{2}+2\eta\Delta_{T}+2\Delta_{T}^{2})}{(2\eta-\Delta_{C})(2\eta-\Delta_{T})}
\end{equation}
\begin{equation}
    \alpha_{21}/\alpha_{11} = \frac{\Delta_{T}^{2}(\eta-2\Delta_{C})+\eta(2\eta^{2}-2\eta\Delta_{C}-\Delta_{C}^{2})+\Delta_{T}(-3\eta^{2}+2\eta\Delta_{C}+2\Delta_{C}^{2})}{(2\eta-\Delta_{C})(2\eta-\Delta_{T})}
\end{equation}
\begin{equation}
    \alpha_{22}/\alpha_{11} =\frac{(\eta-2\Delta_{C})(\eta-2\Delta_{T})}{(2\eta-\Delta_{C})(2\eta-\Delta_{T})}
\end{equation}

The expressions are complicated but we now point out some important features. In typical systems the drive strengths $\Omega_{i}$ are larger than the coupling strength $J$ and as a result $\alpha_{ij}^{(3)} > \alpha_{ij}^{(2)}$. For example, the CZ gate described in the main text is performed with $\Omega_{\text{c}}, \Omega_{\text{t}} \approx 11~\text{MHz}$ on a pair of transmons with estimated coupling $J=2.7~\text{MHz}$. Therefore, the static cross-Kerr can in principle be cancelled by the driven cross-Kerr. The result also gives the leading order linear dependence of the rates $\alpha_{ij}$ on the drive strengths $\Omega_{i}$ and the coupling $J$ as well as their sinusoidal dependence on the relative drive phase $\varphi$ (see Figure 2 in the main text). 

\section{The Weyl and Gell-Mann bases}

When analyzing qubits, we use tensor products of the single-qubit Pauli operators $\bb{P} = \{I, X, Y, Z\}$.
These operators have the following helpful properties:
\begin{enumerate}
    \item they form a projective group under matrix multiplication;
    \item they are unitary;
    \item they are a trace-orthogonal basis for the space of operators; and
    \item they correspond to natural Hamiltonians.
\end{enumerate}
Unfortunately, no set of operators with the same properties exist for higher dimensions.
We thus need to separate some of the properties when analyzing qudits.

There are two sets of operators, namely, the Weyl and Gell-Mann operators that, taken together, satisfy all four properties and also coincide with the Pauli operators in 2D.
We now review these two sets of operators.

The Gell-Mann operators are obtained by embedding the familiar Pauli matrices into two-dimensional subspaces of a higher dimensional space.
Recall that the standard single-qubit Pauli operators are
\begin{align}
    X = \ketbra{1}{0} + \ketbra{0}{1} \notag\\
    Y = i\ketbra{1}{0} - i \ketbra{0}{1} \notag\\
    Z = \ketbra{0} - \ketbra{1}.
\end{align}
We can embed these operators in a $d$-dimensional subspace by defining
\begin{align}
    X^{jk} = \ketbra{j}{k} + \ketbra{k}{j} \notag\\
    Y^{jk} = i\ketbra{j}{k} - i \ketbra{k}{j} \notag\\
    Z^{jk} = \ketbra{j} - \ketbra{k}
\end{align}
for $0 \leq j < k < d$.
The $X^{jk}$ and $Y^{jk}$ operators are trace-orthogonal Hermitian operators and also correspond to transitions between two levels, which is the natural way to control a harmonic oscillator.
Specifically, we can drive the $X^{01}$ and $X^{12}$ Hamiltonians by applying tones at $\omega_{01}$ and $\omega_{12}$, which generate Rabi oscillations in the corresponding qubit subspaces of the qutrit, from which we can generate the single qutrit gates
\begin{equation}
    X_{\pi/2}^{01} = \frac{1}{\sqrt{2}}\begin{pmatrix}
    1 & -i & 0\\
    -i & 1 & 0\\
    0 & 0 & 1
    \end{pmatrix}, \indent     X_{\pi/2}^{12} = \frac{1}{\sqrt{2}}\begin{pmatrix}
    1 & 0 & 0\\
    0 & 1 & -i\\
    0 & -i & 1
    \end{pmatrix}.
\end{equation}
Similarly, we can perform virtual Z gates within these two qubit subspaces of the qutrit, which natively yield the continuous gates
\begin{equation}
    Z^{01}(\phi)  = \begin{pmatrix}
    e^{-i \phi} & 0 & 0\\
    0 & 1 & 0 \\
    0 & 0 & 1
    \end{pmatrix}, \indent     Z^{12}(\phi)  = \begin{pmatrix}
    1 & 0 & 0\\
    0 & 1 & 0 \\
    0 & 0 & e^{i \phi}
    \end{pmatrix}
\end{equation}
From these four gates, $X_{\pi/2}^{01}, X_{\pi/2}^{12},Z^{01}(\phi),Z^{12}(\phi)$, we can compile an arbitrary unitary in $\bb{U}(3)$ using at most 6 of these gates, with a decomposition given in Ref~\cite{PhysRevLett.126.210504}.

However, while the $Z^{jk}$ are Hermitian operators, they are not linearly independent and so do not provide a suitable basis for quantum process tomography.
We thus extend the set $\{X^{jk}, Y^{jk} : 0 \leq j < k < d\}$ to a trace-orthogonal basis by adding a trace-orthogonal set of diagonal operators.
A natural choice would be the projectors onto the computational basis, however, this obscures the fact that all density operators have unit trace.
We thus use the operators
\begin{align}
    D_j = -j \ketbra{j} + \sum_{0 \leq k < j} \ketbra{k}
\end{align}
for $1 \leq j < d$, together with the identity operator to obtain the Gell-Mann basis.
Here the Gell-Mann matrices $\lambda_i$ plus the identity I$_3$ span SU(3) and are a natural choice for qutrit Pauli transfer matrices (PTMs).
For convenience, we index these elements for a single qutrit as follows:
\begin{equation}
    \begin{gathered}
    \textnormal{I}_3 = \begin{pmatrix}
    1&0&0\\
    0&1&0\\
    0&0&1
    \end{pmatrix},
        \lambda_1 = \begin{pmatrix}
    0&1&0\\
    1&0&0\\
    0&0&0
    \end{pmatrix},
        \lambda_2 = \begin{pmatrix}
    0&-i&0\\
    i&0&0\\
    0&0&0
    \end{pmatrix},
        \lambda_3 = \begin{pmatrix}
    1&0&0\\
    0&-1&0\\
    0&0&0
    \end{pmatrix},
        \lambda_4 = \begin{pmatrix}
    0&0&1\\
    0&0&0\\
    1&0&0
    \end{pmatrix}, \\
        \lambda_5 = \begin{pmatrix}
    0&0&-i\\
    0&0&0\\
    i&0&0
    \end{pmatrix},
        \lambda_6 = \begin{pmatrix}
    0&0&0\\
    0&0&1\\
    0&1&0
    \end{pmatrix},
        \lambda_7 = \begin{pmatrix}
    0&0&0\\
    0&0&-i\\
    0&i&0
    \end{pmatrix},
        \lambda_8 = \frac{1}{\sqrt{3}}\begin{pmatrix}
    1&0&0\\
    0&1&0\\
    0&0&-2
    \end{pmatrix}
    \end{gathered}
\end{equation}

Having defined the Gell-Mann basis, we now define the Heisenberg-Weyl operators, which are a unitary generalization of the familiar single-qubit Pauli operators to higher dimensional spaces (qudits) that enable the Clifford group to be generalized.
Let $d$ be a positive integer and $\bb{Z}_d = \{0, \ldots, d - 1\}$ denote the set of integers modulo $d$.
The generalizations of the $X$ and $Z$ operators to qudits are
\begin{align}
    X &= \sum_{j \in \bb{Z}_d} \ketbra{j \oplus_d 1}{j} \notag\\
    Z &= \sum_{j \in \bb{Z}_d} \exp\left(\frac{2 \pi i }{d}j\right)\ketbra{j}{j},
\end{align}
where $\oplus_d$ denotes addition modulo $d$. From this, the definition of the two-qudit generalization of a controlled-Z gate (CZ) follows naturally as:
\begin{equation}
    U_{CZ} = \sum_{n =1}^{d} \ketbra{n} \otimes Z^n
\end{equation}
For the two qutrit entangling gates performed in this work, this yields the follow unitaries:
\begin{equation}
   U_{CZ} = e^{2\pi i /3}( \ketbra{11}{11} + \ketbra{22}{22}) + e^{4\pi i /3}( \ketbra{12}{12} + \ketbra{21}{21}) + \sum_{j=0}^2 (\ketbra{0j}{0j} + \ketbra{j0}{j0})  
\end{equation}
\begin{equation}
    U_{CZ^\dag} = e^{4\pi i /3}( \ketbra{11}{11} + \ketbra{22}{22}) + e^{2\pi i /3}( \ketbra{12}{12} + \ketbra{21}{21}) + \sum_{j=0}^2 (\ketbra{0j}{0j} + \ketbra{j0}{j0})  
\end{equation}
The Weyl basis for $\bb{C}^{d \times d}$ is the set
\begin{align}
    \bb{W}_d = \{W_{xz} = X^x Z^z : x,z \in \bb{Z}_d \}\,,
\end{align}
which is a trace-orthogonal basis for $\bb{C}^{d \times d}$ as it satisfies
\begin{align}
    \tr W^\dagger V = d \delta_{W, V} \quad \forall\,W, V \in \bb{W}_d.
\end{align}
Let $n$ be a positive integer and $D = d^n$. 
Then we define the $n$-qudit Weyl basis to be the set $\bb{W}_{d, n} = \bb{W}_d^{\otimes n}$, which is a trace-orthogonal basis for $\bb{C}^{D \times D}$.
Note that the Weyl basis is \emph{not} a proper group as it is not closed under multiplication.
However, every element of the closure of the Weyl basis is proportional to an element of the Weyl basis up to an overall phase.
This overall phase vanishes in all cases as we only consider the adjoint action of Weyl operators and so we treat the Weyl basis as a projective group.
The $n$-qutrit Clifford group is then defined to be the normalizer of the extended Weyl group $\bb{EW}_{d, n} = \bb{U}(1) \bb{W}_{d, n}$ (which is a proper group), that is,
\begin{align}
    \mathfrak{C}_{d, n} = \{ U \in \bb{U}(d^n) : U \bb{EW}_{d, n} U^\dagger = \bb{EW}_{d, n} \}.
\end{align}
\newpage
\section{Assumptions for characterization}

We now define the assumption of time-dependent Markovian noise that we use to analyze the results of our characterization protocols.
At a high level, we assume that each operation applied to the system corresponds to a linear map that is independent of what other maps have been applied but may depend on the number of operations that have been applied since the system was initialized.
We allow this time dependence primarily to facilitate post-processing of time-stationary noise processes wherein the physical process does not depend on the number of operations that have been applied since the system was initialized but we are performing a weighted average over the operation that is applied in the $j$th time step.

Formally, for a vector space $\bb{V}$ let $\bb{V}^*$ denote the dual of $\bb{V}$ and $\mc{L}(\bb{V})$ denote the set of linear maps from $\bb{V}$ to itself.
Then we assume the following.
\begin{enumerate}
    \item The state space of a physical implementation of an $n$-qudit system is some fixed vector space $\bb{V}$.
    \item Preparing the system in the state $\rho$ corresponds to setting the state of the quantum system to some $\Theta_\rho \in \bb{V}$.
    \item Applying some unitary operation $U$ to the system in the $j$th time step after preparing the system in a state corresponds to applying some linear map, $\Theta(j, U) \in \mc{L}(\bb{V})$ to the state of the system.
    \item The expectation value of an observable $Q$ is obtained by applying some fixed $\Theta_Q \in \bb{V}^*$. (Note that we will ignore finite measurement statistics for now.)
\end{enumerate}
With some abuse of notation, we refer to the functions $\Theta_j : \bb{N} \times \bb{U}(d^n) \to \mc{L}(\bb{V} \to \bb{V})$ and the vectors $\Theta_\rho$ and $\Theta_Q$ together as the implementation map $\Theta$~\cite{Helsen2020AGF}.

The assumption of time-dependent Markovian noise allows hidden Markovianity as the implementation map can include a coupling to an environment.
This hidden Markovianity is frequently referred to as non-Markovianity in the quantum information community.
However, we allow it in the general setting because the additional assumptions (such as positivity and complete positivity) are more cumbersome to define and typically are only helpful in the final steps of an analysis.
Indeed, it is conceptually useful to include post-processing steps that depend only on one time step into an ``effective'' implementation map.

An ideal isolated implementation is one wherein there is a linear isomorphism between $\bb{V}$ and $\bb{C}^{D \times D}$.
For concreteness, we define an isomorphism $\dket{*}_\bb{B} : \bb{C}^{D \times D} \to \bb{V}$ relative to a trace-orthonormal basis $\bb{B} \subset \bb{C}^{D \times D}$ as follows.
As $\bb{B}$ is a trace-orthonormal basis, we can write any $A \in \bb{C}^{D \times D}$ in terms of $\bb{B}$ as
\begin{align}
    A = \sum_{B \in \bb{B}} \tr(B^\dagger A) B.
\end{align}
Therefore we can choose $\{ \dket{B}_\bb{B} : B \in \bb{B}\}$ to be an orthonormal basis of $\bb{V}$ and extend it to an isomorphism by defining
\begin{align}
    \dket{A}_\bb{B} = \sum_{B \in \bb{B}} \tr(B^\dagger A) \dket{B}_\bb{B}.
\end{align}
We will typically suppress the subscript $\bb{B}$ as it will be clear from the context (either the normalized Weyl basis or the normalized Gell-Mann basis).
To avoid having to define normalized versions of the bases explicitly, we define
\begin{align}
    \hat{B} = B / \sqrt{\tr B^\dagger B}.
\end{align}
Moreover, defining $\dbra{A} = \dket{A}^\dagger$ and $\dbraket{A}{B} = \dbra{A} \dket{B}$ and using the fact that $\bb{B}$ is a trace-orthonormal basis and $\{\dket{B}: B \in \bb{B}\}$ is an orthonormal basis, we have
\begin{align}
    \tr A^\dagger T 
    &= \sum_{B, C \in \bb{B}} \tr(A^\dagger B) \tr(C^\dagger T) \tr(B^\dagger C) \notag\\
    &= \sum_{B \in \bb{B}} \tr(A^\dagger B) \tr(B^\dagger T) \notag\\
    &= \dbraket{A}{T}.
\end{align}

With the above isomorphism, we can define the ideal implementation of a unitary operator $U \in \bb{U}(D)$ to be
\begin{align}
    \phi_\bb{B}(U) = \sum_{B \in \bb{B}} \dketbra{U B U^\dagger}{B},
\end{align}
which, by linearity, will satisfy
\begin{align}
    \phi_\bb{B}(U) \dket{A} = \dket{U A U^\dagger}.
\end{align}
As above, we will also suppress the $_\bb{B}$ on $\phi$ when the basis is clear from the context.
Moreover, as $U \bb{B} U^\dagger$ is a trace-orthonormal basis, one can readily verify that $\phi(U) \phi(V) = \phi(UV)$ for all $U, V \in \bb{U}(D)$, that is, $\phi$ is a representation of $\bb{U}(D)$.

\newpage
\section{Cross Entropy Benchmarking}

\begin{figure}[h!]
    \centering
    \includegraphics[width =\textwidth]{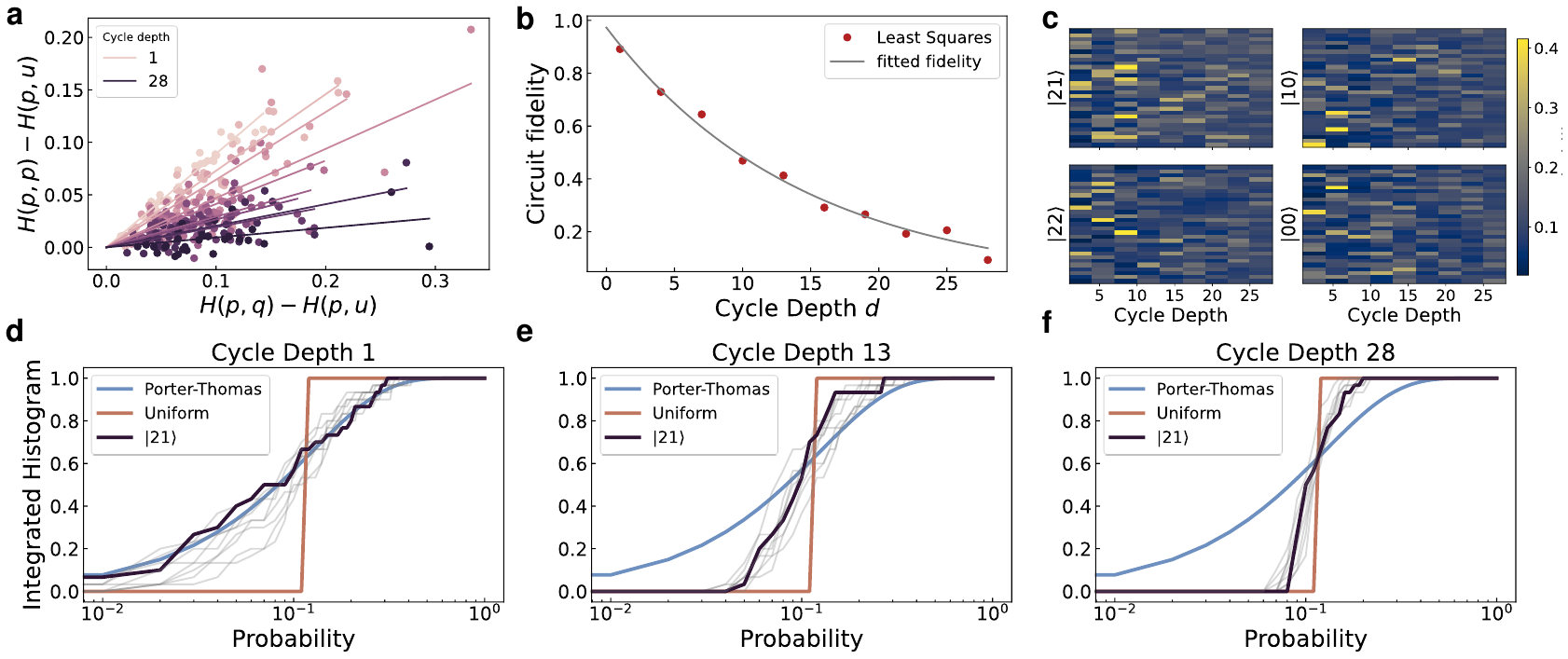}
    \caption{\textbf{Cross entropy benchmarking (XEB) for qutrit}, \textbf{a}, For the
    XEB circuits, the entropy difference $H(p,q) - H(p,u)$ is plotted against the ideal $H(p,p) - H(p,u)$ at each cycle depth, with 30 randomizations at each depth. The linear fit gives the fidelity at that particular depth. \textbf{b}, We show the exponential decay of the depolarized CZ$^\dag$ fidelity obtained from the linear fits in \textnormal{a}. \textbf{c}, We plot an example of the speckle purity decay as a function of cycle depth for a representative set of four states in the two-qutrit Hilbert space. The probabilities of measuring the given tritstring are shown as a function of cycle depth across all 30 randomizations. The bright ``speckle'' pattern characteristic of the Porter-Thomas distribution at low circuit depths is smoothed out at larger depths. \textbf{d}-\textbf{f}, We show that the distribution of tritstrings transitions from the Porter-Thomas ditribution at low depth, to a uniform distribution at deeper depths. The CDF is emphasized for a representative tritstring, $\ket{21}$, while the CDFs of the other tritstrings are shown in grey to show typical variation of the CDF across different states.}
    \label{fig:XEB}
\end{figure}

As discussed in the main text, performing randomized benchmarking (RB) on a two qutrit gate is prohibitively expensive \cite{PhysRevLett.126.210504}. We utilize cross entropy benchmarking (XEB) as a second SPAM (state preparation and measurement) free benchmarking protocol to corroborate the fidelity obtained from cycle benchmarking (CB). Recently XEB played a central role in the quantum supremacy experiment in \cite{arute2019quantum} and was used to benchmark the non-Clifford, three qubit $i$Toffoli gate in \cite{Kim2022}. XEB theory has been discussed in detail in previous works so we present only a brief review of the method before outlining more explicitly how it can be used to benchmark the two-qutrit CZ gate demonstrated here.

For XEB purposes a convenient definition of a random quantum circuit (RQC) is a quantum circuit randomly selected from an ensemble of circuits such that the distribution of probabilities (across the ensemble of circuits) of observing a particular ditstring, $x$, follows the Porter-Thomas distribution (for all possible ditstrings $x$) \cite{MullaneXEBTheory}. In the present case of quantum circuits involving two qutrits we consider the probabilities of tritstrings 00, 01, \dots, 21, 22. After averaging over an ensemble of random circuits errors are tailored to be purely depolarizing so an error corresponds to the circuit outputting a fully mixed state. In a mixed state the tritstring distribution is uniform with each outcome $x_{i}$ having equal probability of $1/3^{n}$ for an $n$ qutrit system. Thus, intuitively the circuit fidelity (under this error model) can be thought of as the deviation of the measured tritstring distribution from the uniform distribution. The XEB fidelity makes this relationship precise as we now outline. 

Denote the possible tritstrings $x_{i}$ for $i=1,\dots,3^{n}$ and let $p(x_{i})$ be the ideal tritstring distribution for the output of a particular quantum circuit. Then $q(x_{i})$ is the measured distribution. The linear cross entropy of two probability distributions $p_{1}(x)$ and $p_{2}(x)$ with the same support is defined as
\begin{equation}
    H(p_{1},p_{2}) = \sum_{x}p_{1}(x)p_{2}(x)
\end{equation}
where the sum runs over the full support of the probability distributions and the self entropy is written compactly as $H(p_{1})\equiv H(p_{1},p_{1})$. Under the depolarizing error model it can be shown straightforwardly that the circuit fidelity is
\begin{equation}
    F_{\text{XEB}}=\frac{H(p,q)-H(p,u)}{H(p,p)-H(p,u)}
\end{equation}
where $u(x_{i})$ is the uniform probability distribution \cite{arute2019quantum}. This is ultimately the difference in the ideal to measured and ideal to uniform cross entropies, normalized by the difference if the measured distribution were to perfectly match the ideal distribution ($p(x_{i})=q(x_{i})$ for all $i$). 

We follow the XEB protocol for benchmarking gates/cycles outlined in Ref.~\cite{arute2019quantum}. This consists of measuring two qutrit circuits of varying cycle depths $M$ where cycle $i$ consists of two single qutrit gates $G_{i,j}$ where $i=1,\dots,M$ labels the cycle and $j=1,2$ labels the qutrit, followed by the two qutrit CZ gate$^\dag$ (see Figure \,\ref{fig:XEB}a). The single qutrit gates are randomly selected unitaries from $SU(3)$ and decomposed according to the decomposition given in \cite{Dita_2003}. For each cycle depth $M$ we generate $N$ random circuits to be run and compute the ideal probability distributions of the output tritstrings for each circuit. The XEB fidelity at cycle depth $M$, $F_{\text{XEB},M}$ is determined by performing a least squares fit of the linear relationship between $H(p_{i},q_{i})-H(p_{i},u)$ and $H(p_{i},p_{i})-H(p_{i},u)$ with $i=1,\dots,N$ labeling the random circuit at the given depth (see Figure). The cycle infidelity, $\epsilon_{\text{cycle}}$, is estimated from the exponential decay of $F_{\text{XEB},M}$ as a function of cycle depth $M$. The error rate extracted in this way agrees well with CB (see Benchmarking section in the main text).

Next we validate our implementation of XEB by verifying that the distribution of probabilities of a given tritstring across the ensemble of random circuits does indeed approach the Porter-Thomas distribution,
\begin{equation}
    \mathcal{P}(p)=(D-1)(1-p)^{D-2} \approx De^{-Dp}
\end{equation}
where again $D$ is the Hilbert space dimension, $D=3^{2}$ for two qutrits and the approximate equality holds in the limit of large Hilbert space dimension, $D \gg 1$. Next we consider the evolution of this distribution as a function of cycle depth. At short to intermediate cycle depths, the circuit is sufficiently random and the distribution approaches Porter-Thomas. At larger cycle depths, the distribution begins to converge to $\mathcal{P}(p) \rightarrow \delta(p-1/D)$ since the depolarizing errors dominate and the tritstring distributions approach the uniform distribution for all circuits. We plot our experimental observation of this behavior in Fig.\,\ref{fig:XEB}d-f.

The method of Speckle Purity Benchmarking (SPB) is based on this observation. Denoting the state purity as $\gamma$, it can be estimated at depth $M$ from the raw results of the XEB protocol outlined above by the relationship
\begin{equation}
    \gamma(M)=\text{Var}(p_{M})\frac{D^{2}(D+1)}{(D-1)}
\end{equation}
where $p_{M}$ is the set of measured probabilities of a given tritstring $x$ across the $N$ random circuits at cycle depth $M$ \cite{arute2019quantum}. Thus, once we have demonstrated that the distribution does indeed converge to the Porter-Thomas distribution we can estimate the decay of the state purity from just the variance of the distribution, without the need for state tomography which requires an exponential number of measurements, $3^{n(d-1)}$ where $n$ is the number of qudits and $d$ the dimension of each qudit. For our present case of two qutrits we would need to perform 81 measurements per circuit to determine the state purity by state tomography.
\newpage
\section{Extracting an Error Budget Estimation from Cycle Benchmarking Results}

\begin{figure}[h!]
    \centering
    \includegraphics[width = \textwidth]{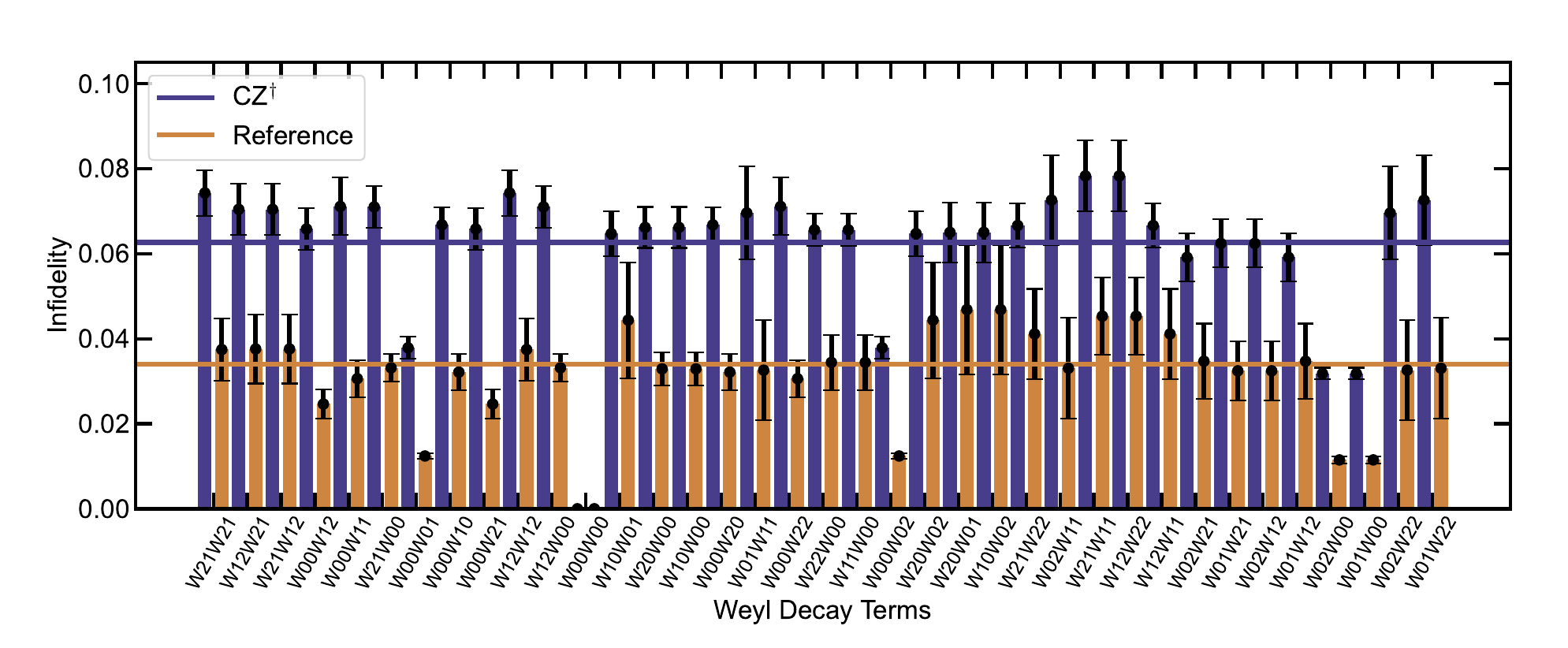}
    \caption{\textbf{Cycle benchmarking results with Weyl decay terms of CZ$^\dag$ gate}. The cycle benchmarking results shown analyze a subset of 53 Weyl channels at depths $m \in \{0,3,6,15 \}$. In order to provide a SPAM free estimation of the error, we compare our CB results to a reference CB experiment without the two qutrit gate to isolate and subtract the error due to the Weyl twirling itself. Here we plot only the shared decay channels analyzed for both the CZ$^\dag$ and reference CB experiments. }
    \label{fig:exp_CB_with_weyls}
\end{figure}

We now outline how data from cycle benchmarking~\cite{cb} can be used to break down the error budget for a multi-qudit Clifford gate.
The following analysis also generalizes that of \cite{cb} to gate-dependent noise and allows some non-Markovian errors (namely, couplings to an environment).

\subsection{The cycle benchmarking protocol}

We now outline the cycle benchmarking protocol.
Let $d$, $m$, and $n$ be positive integers, $\psi$ be a quantum state, and $Q$ be an observable.
Then a cycle benchmarking circuit consists of the following:
\begin{enumerate}
    \item Prepare the system in a state $\psi$;
    \item For each $j \in \bb{Z}_m$,
    \begin{enumerate}
        \item Choose an $n$-qudit Weyl operator $W_j \in \bb{W}_d^{\otimes n}$ uniformly at random;
        \item Apply $W_j$ to the system; and
        \item Apply $C$ to the system.
    \end{enumerate}
    \item Choose an $n$-qudit Weyl operator $W_m \in \bb{W}_d^{\otimes n}$ uniformly at random;
    \item Apply $W_m$ to the system; and
    \item Measure the expectation value of an operator $Q$.
\end{enumerate}

For time-dependent Markovian noise, the expectation value for the cycle benchmarking protocol for a fixed choice of random Weyl operators $\vec{W} = (W_0, \ldots , W_m)$, preparation $\psi$ and operator $Q$ relative to an implementation map $\Theta$ is
\begin{align}\label{eq:cb}
    \mu_{\Theta, \psi, \vec{W}, Q} = \Theta_Q \left(\prod_{j = m \to 1} \Theta(2j, W_j) \Theta(2j - 1, C) \right)\Theta(0, W_0) \Theta_\psi.
\end{align}
The Weyl operators are sampled uniformly and independently and are uncorrelated in \cref{eq:cb}.
The average over all choices of random Weyl operators is then
\begin{align}
    \bb{E}_{\vec{W}} \mu_{\Theta, \psi, \vec{W}, Q} = \Theta_Q \left(\prod_{j = m \to 1} \Theta(j) \Theta(2j - 1, C) \right) \Theta(0)\Theta_\psi \,,
\end{align}
where for an implementation map $\Theta$ we define
\begin{align}
    \Theta(j) &= \bb{E}_{W \in \bb{W}_{d, n}} \Theta(2j, W).
\end{align}
That is, we have factorized the average of the expectation values into a product of independent averages.

\subsection{CB decay}

Before turning to analyze the variance, we first show how the above method is equivalent to that of Ref.~\cite{cb} under equivalent assumptions.
First, let $\Lambda$ denote the physical implementation map and assume that for all $W \in \bb{W}_{d, n}$, we have
\begin{align}
    \Lambda(W) = \mc{L} \phi(W) \mc{R}
\end{align}
for some linear maps $\mc{L}$ and $\mc{R}$.
We then redefine
\begin{align}
\Lambda_\psi &\to \mc{R} \Lambda_\psi \notag\\
\Lambda(2j - 1, C) &\to \mc{R} \Lambda(2j - 1, C) \mc{L}, \notag\\
\Lambda_Q &\to \Lambda_Q \mc{L},
\end{align}
so that without loss of generality we can set $\mc{L}$ and $\mc{R}$ to the identity, that is, we can assume that the implementation of the random operations is effectively ideal.

One can readily verify that the ideal implementation of a Weyl operator $W$ satisfies
\begin{align}
    \phi(W) = \sum_{V \in \bb{W}_{d, n}} \dketbra{W \hat{V} W^\dagger}{\hat{V}} = \sum_{V \in \bb{W}_{d, n}} \chi_V(W) \dketbra{\hat{V}}{\hat{V}}
\end{align}
where
\begin{align}
    \chi_V(W) = \dbraket{\hat{V}}{W \hat{V} W^\dagger} = D^{-1} \tr V^\dagger W V W^\dagger,
\end{align}
which is a character of the projective Weyl group.
By Schur's orthogonality relations, for any $U \in \bb{W}_{d, n}$ we have
\begin{align}
    \bb{E}_{W \in \bb{W}_{d, n}} \sum_{V \in \bb{W}_{d, n}} \chi_U^*(W) \chi_V(W) \dketbra{\hat{V}}{\hat{V}} = \dketbra{\hat{U}},
\end{align}
so that if we set $\Theta(2j, W_j) = \chi_{U_j}^*(W_j) \phi(W_j)$ for $U_j \in \bb{W}_{d, n}$, which can be accomplished by multiplying $\mu_{\Theta, \vec{W}}$ by $\chi_{U_j}^*(W_j)$ for each $j$, we find that \cref{eq:cb} simplifies to
\begin{align}\label{eq:cb_scalar}
    \Lambda_Q \dketbra{\hat{U}_m}{\hat{U}_0} \Lambda_\psi \prod_{j = m \to 1} \dbra{\hat{U}_j} \Theta(2j - 1, C) \dket{\hat{U}_{j-1}},
\end{align}
which is now a product of scalars so we can reorder terms as desired.
When $C$ is a Clifford operator, choosing $U_j = C^j U_0 C^{-j}$ for some fixed $U_0 \in \bb{W}_{d, n}$ reduces \cref{eq:cb_scalar} to the expression in Ref.~\cite{cb}.

\subsection{CB variance}

We now analyze the variance to lowest order to show how the unitarity~\cite{Wallman2015} can be estimated.
The analysis closely parallels that of Ref.~\cite{Wallman2015}, except that because we are using a weaker twirl (namely, over the Heisenberg-Weyl group instead of over the full multi-qudit Clifford group), the matrix that governs the decay rates has more eigenvalues.
As in Ref.~\cite{Wallman2015}, We use the fact that for any scalar $\lambda$, we have $|\lambda|^2 = \lambda \otimes \lambda^*$ and that multiplication distributes across tensor products, so that we will consider the implementation map $\Theta = \Lambda \otimes \Lambda^*$, where the phases added when analyzing a decay cancel.
From \cref{eq:cb}, we then have
\begin{align}\label{eq:meanSq}
    \bb{E}_{\vec{W}} |\mu_{\Lambda, \psi, \vec{W}, Q}|^2
    &= \bb{E}_{\vec{W}} \mu_{\Theta, \psi, \vec{W}, Q} \notag\\
    &= \Theta_Q \left(\prod_{j = m \to 1} \Theta(j) \Theta(C) \right) \Theta(0) \Theta_\psi.
\end{align}
Applying Schur's orthogonality relations gives
\begin{align}
    \Theta(j) 
    &= \bb{E}_{W \in \bb{W}_{d, n}} \phi(W) \otimes \phi^*(W) \notag\\
    &= \sum_{U, V \in \bb{W}_{d, n}} \dketbra{\hat{U} \otimes \hat{V}} \bb{E}_{W \in \bb{W}_d^{\otimes n}} \chi_U(W) \chi^*_V(W) \notag\\
    &= \sum_{U, V \in \bb{W}_{d, n}} \dketbra{\hat{U} \otimes \hat{V}} \delta_{U, V} \notag\\
    &= \sum_{U \in \bb{W}_{d, n}} \dketbra{\hat{U} \otimes \hat{U}}.
\end{align}
In particular, $\Theta(j)^2 = \Theta(0)$ for all $j$ and so \cref{eq:meanSq} can be written as
\begin{align}\label{eq:unitarity_decay}
    \bb{E}_{\vec{W}} |\mu_{\Lambda, \psi, \vec{W}, Q}|^2 = \Theta_Q M^m \Theta_\psi,
\end{align}
where
\begin{align}
    M = \Theta(0) \Theta(C) \Theta(0) = \sum_{U, V \in \bb{W}_{d, n}} \dketbra{\hat{U} \otimes \hat{U}}{\hat{V} \otimes \hat{V}} |\Lambda_{UV}(C)|^2.
\end{align}
For simplicity, we assume that $\Lambda(C)$ is unital and trace preserving, that is, that for all $U \in \bb{W}_{d, n}$ we have
\begin{align}
    \Lambda_{U, I}(C) = \Lambda_{I, U}(C) = \delta_{I, U}.
\end{align}
Then we can rewrite \cref{eq:unitarity_decay} as
\begin{align}\label{eq:unitarity_decay2}
    \bb{E}_{\vec{W}} |\mu_{\Lambda, \psi, \vec{W}, Q}|^2 
    &= \Theta_Q M_{\rm u}^m \Theta_\psi + \Theta_Q \dketbra{\hat{I}^{\otimes 2}} \Theta_\psi \notag\\
    &= \Theta_Q M_{\rm u}^m \Theta_\psi + |\Lambda_Q \dketbra{\hat{I}} \Lambda_\psi|^2,
\end{align}
where we define the unital block of $M$ to be
\begin{align}
    M_{\rm u} = \Theta(0) \Theta(C) \Theta(0) = \sum_{U, V \in \bb{W}^*_{d, n}} \dketbra{\hat{U} \otimes \hat{U}}{\hat{V} \otimes \hat{V}} |\Lambda_{UV}(C)|^2
\end{align}
and $\bb{W}^*_{d, n} = \bb{W}_{d, n} - \{ I \}$.
Further, note that the constant term is can be directly estimated from \cref{eq:cb_scalar} for any $C$ by setting $U_0 = I$, and so for unital noise the variance over random Weyl operators satisfies
\begin{align}\label{eq:unitarity_decay3}
    \bb{E}_{\vec{W}} |\mu_{\Lambda, \psi, \vec{W}, Q}|^2 - |\bb{E}_{\vec{W}} \mu_{\Lambda, \psi, \vec{W}, Q}|^2
    &= \Theta_Q M_{\rm u}^m \Theta_\psi.
\end{align}

We now show how the unitarity of a channel~\cite{Wallman2015} can be estimated from \cref{eq:unitarity_decay3} under assumptions similar to those in XEB.
First, note that if $M_{\rm u}$ is diagonalizable, \cref{eq:unitarity_decay} can be written as
\begin{align}
    \bb{E}_{\vec{W}} |\mu_{\Lambda, \vec{W}}|^2 = \sum_j \alpha_j \lambda_j^m
\end{align}
where the $\lambda_j$ are the nonzero eigenvalues of $M_{\rm u}$ and the $\alpha_j$ are the overlaps of the corresponding eigenvectors of $M_{\rm u}$ with the SPAM vectors.
We thus want to prove that the unitarity of a channel corresponds to an eigenvalue of $M_{\rm u}$.
The unitarity $u$ of a quantum channel is defined to be~\cite{Wallman2015}
\begin{align}
    u^2 = \frac{1}{D^2 - 1} \sum_{U, V \in \bb{W}^*_{d, n}} |\Lambda_{UV}(C)|^2.
\end{align}
As in XEB, we now assume that the error model consists only of a unitary error and global depolarization, so that
\begin{align}
    \Lambda(C) = \phi(T) \mc{D}_p
\end{align}
for some unknown $T \in \bb{U}(D)$ where
\begin{align}
    \mc{D}_p(\rho) = p \rho + (1 - p) I/D
\end{align}
is the global depolarizing channel.
Under this assumption, we have $u = p$, so that the goal is to learn an effective depolarizing rate of the noise process.
We now show that
\begin{align}
    v = \sum_{U \in \bb{W}^*_{d, n}} \dket{\hat{U}^{\otimes 2}}
\end{align}
is an eigenvector of $M_{\rm u}$ with eigenvalue $p^2$.
First note that for all $U \in \bb{W}^*_{d, n}$, we have
\begin{align}\label{eq:col_norm}
    \sum_{V \in \bb{W}^*_{d, n}} |\Lambda_{U, V}|^2 = p^2.
\end{align}
Therefore
\begin{align}
    M_{\rm u} v 
    &= \sum_{U, V, W \in \bb{W}^*_{d, n}} \dket{\hat{U}^{\otimes 2}} |\Lambda_{U, V}(C)|^2 \dbraket{\hat{V}^{\otimes 2}}{\hat{W}^{\otimes 2}} \notag\\
    &= \sum_{U \in \bb{W}^*_{d, n}} \dket{\hat{U}^{\otimes 2}} p^2 \notag\\
    &= p^2 v
\end{align}
that is, we have an eigenvector of $M_{\rm u}$ the purity matrix whose eigenvector is exactly $p^2$.
Moreover, by \cref{eq:col_norm}, $M_{\rm u} / p^2$ is a stochastic matrix and so its largest eigenvalue is 1. 
Therefore, under the assumption of global depolarizing and unitary noise, the largest eigenvalue of the unital part of $M_{\rm u}$ is $u^2$.

We now show how we can estimate the eigenvalues of $M_{\rm u}$.
Let $s$ be the order of $C$, that is, the smallest positive integer such that $C^s \propto I$ and assume that $s$ divides $m$.
Then in the limit of high fidelity, the eigenvectors of $M_{\rm u}^s$ are $\dket{\hat{U}^{\otimes 2}}$, and so we can find the largest eigenvalue of $M_{\rm u}$ (and hence the purity under the assumption of global depolarizing and unitary noise) by finding the slowest decay rate of \cref{eq:unitarity_decay3}.
\newpage
\section{Qutrit CZ Gate Benchmarking}
Here we provide the cycle benchmarking results of the 783 ns qutrit CZ gate from the main text. As with the CZ$^\dag$, we estimate the process fidelity of the isolated CZ gate by comparing our CB results to a reference cycle, and using the formula:
\begin{equation}
    e_F = \frac{D-1}{D}\Big(1 - \frac{\mathcal{F}_{\textnormal{CZ}}}{\mathcal{F_{\textnormal{Reference}}}}\Big)
\end{equation} where $D$ is the dimension of the Hilbert space of one's system. In our benchmarking of the CZ, we find a Weyl infidelity of 0.0861 for the dressed CZ cycle and 0.034 for the reference cycle. Ultimately isolating the errors of the CZ gate, we find an estimated process fidelity of 95.2(3)\%.

` 
\begin{figure}[h!]
    \centering
    \includegraphics[width = \textwidth]{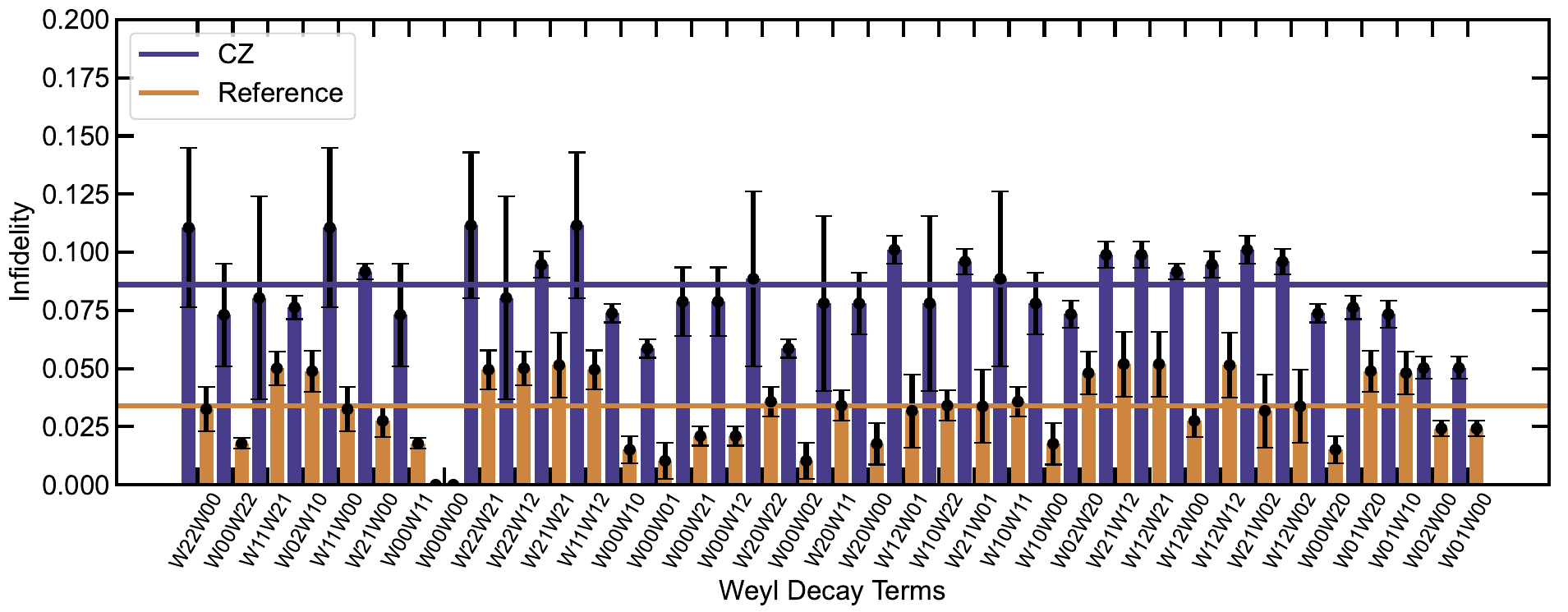}
    \caption{\textbf{Cycle benchmarking results with Weyl decay terms of the qutrit CZ gate}. The cycle benchmarking results analyze a subset of 54 Weyl channels at depths $m \in \{0,3,6 \}$ for the CZ cycle. In order to provide a SPAM free estimation of the error, we again compare our CB results to a reference CB experiment. Here we plot only the shared decay channels analyzed for both the CZ and reference CB experiments. }
    \label{fig:cz_cb_supp}
\end{figure}
\newpage
\section{Frequency dependence of driven cross-Kerr}
\begin{figure}[h!]
    \centering
    \includegraphics[width = 0.9\textwidth]{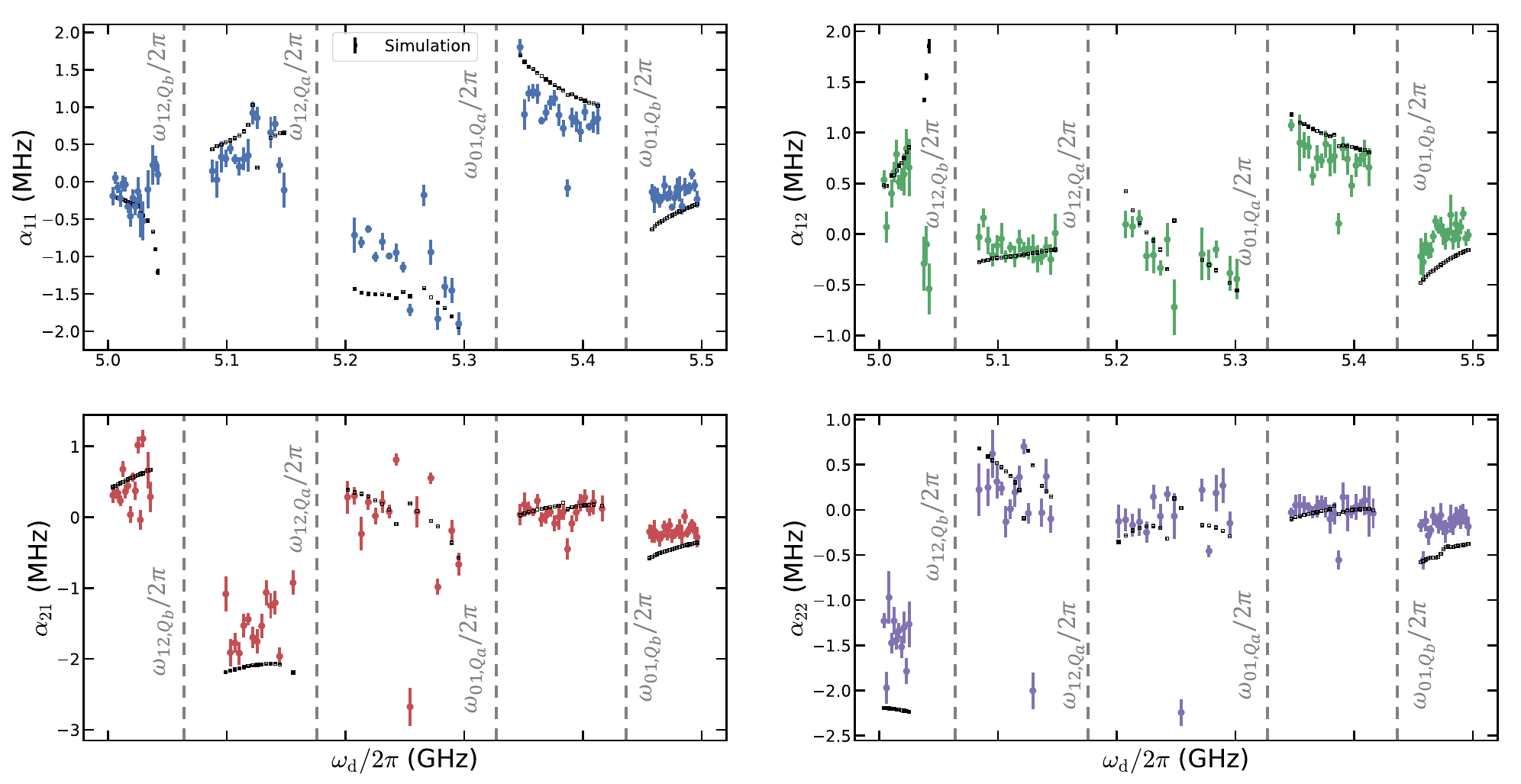}
    \caption{\textbf{Frequency dependence of the full microwave-activated cross-Kerr Hamiltonian}. We compare the dependence of all of the $\alpha_{ij}$ in the driven cross-Kerr Hamiltonian on the frequency of the drive $\omega_d$ using an \textit{ab-initio} master equation simulation in QuTiP.}
    \label{fig:freq_supp}
\end{figure}

In Figure 2 in the main text, we showed the results of fitting the frequency dependence of the $\alpha_{12}$ term in the driven cross-Kerr Hamiltonian to an \textit{ab-initio} master equation simulation. Here we provide some additional details on how this simulation was performed in QuTiP \cite{JOHANSSON20121760,JOHANSSON20131234} and present the results for measuring and characterizing the frequency dependence of all four of the $\alpha_{ij}$ in the driven cross-Kerr Hamiltonian. We note that transient TLS features and higher transitions meant that some of the data did not fit to a linear model, we therefore only plot data where the uncertainty on our linear fit (the source of the error bars) was less than 300 KHz. Similarly, our simulation at times produced unphysical results near transitions, with very large cross-Kerr; in the interest of readability, we therefore also omitted points where the magnitude of the simulated cross-Kerr was larger than 3 MHz.

To perform the master equation simulation, we considered the Hamiltonian of two fixed frequency transmons, with a fixed capacitive coupling from a coplanar waveguide resonator. For the frequencies and anharmonicities of the pair of transmons, we used the experimental parameters from our chip (as can be found in Table \ref{tab:single_qutrit_parameters_cz}). We first found the strength of the capacitive coupling, $J$, by adjusting it until the simulated parameters for the always on, static $\alpha_{ij}$ best matched our experimental measurements of these parameters. We then fixed a single drive frequency, $\omega_d$, and performed a master equation simulation of the simultaneous Stark drives at that frequency until our data best matched the values found for that point in Figure \ref{fig:freq_supp}. After this, we extrapolated those parameters to be the same across all frequencies of the Stark driving, and simulate using them for the rest of the frequencies in Figure \ref{fig:freq_supp}.

\newpage
\section{Quantum Process Tomography}
\begin{figure}[h!]
    \centering
    \includegraphics[width = 1.1\textwidth]{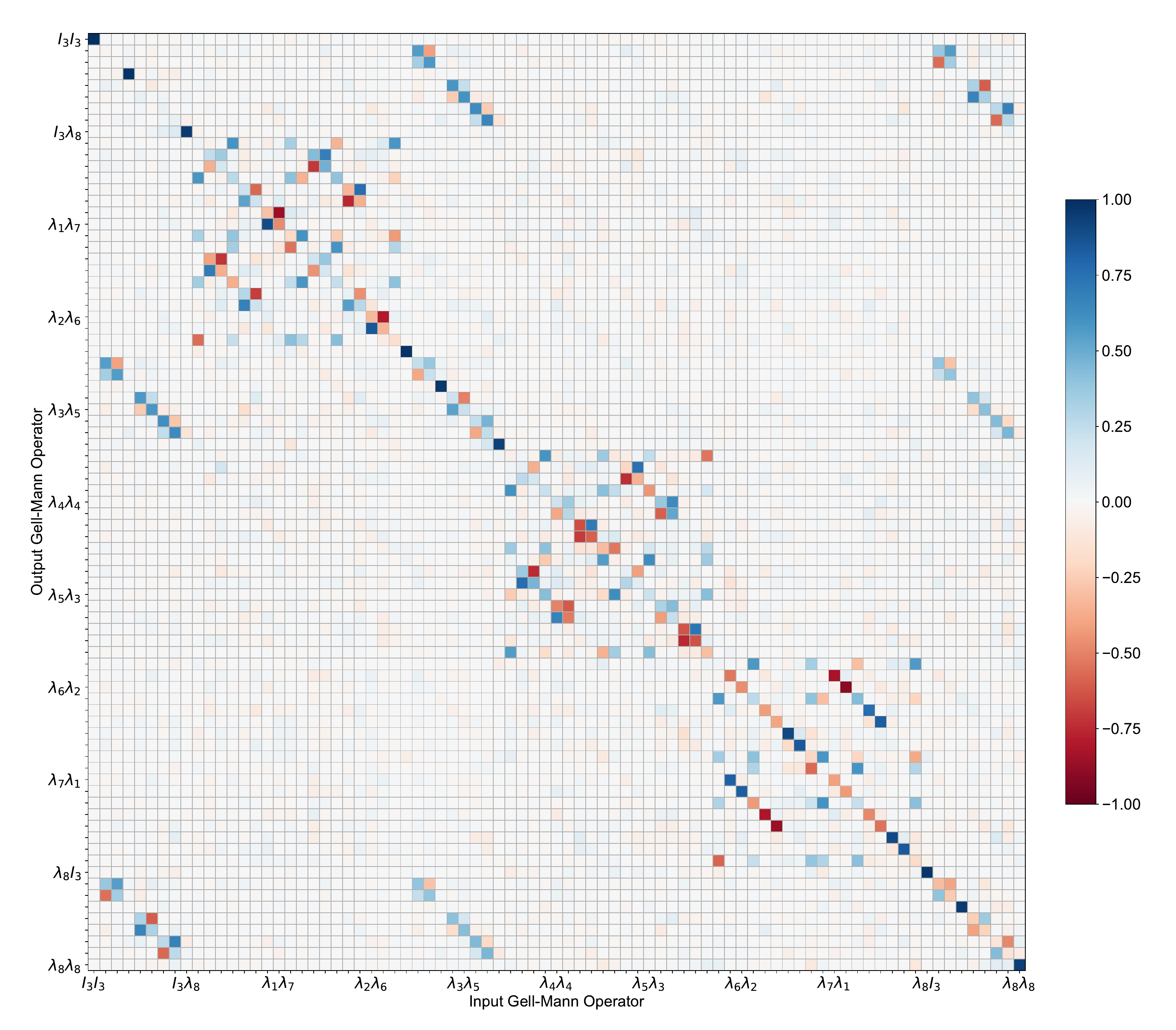}
    \caption{\textbf{CZ$^\dag$ process matrix}. Experimentally reconstructed Process Matrix of Qutrit CZ$^\dag (\mathcal{E}_{\textnormal{exp}}$) gate with process fidelity of 93.2\%.}
    \label{fig:exp_QPT}
\end{figure}

\begin{figure}[h!]
    \centering
    \includegraphics[width = 1.1\textwidth]{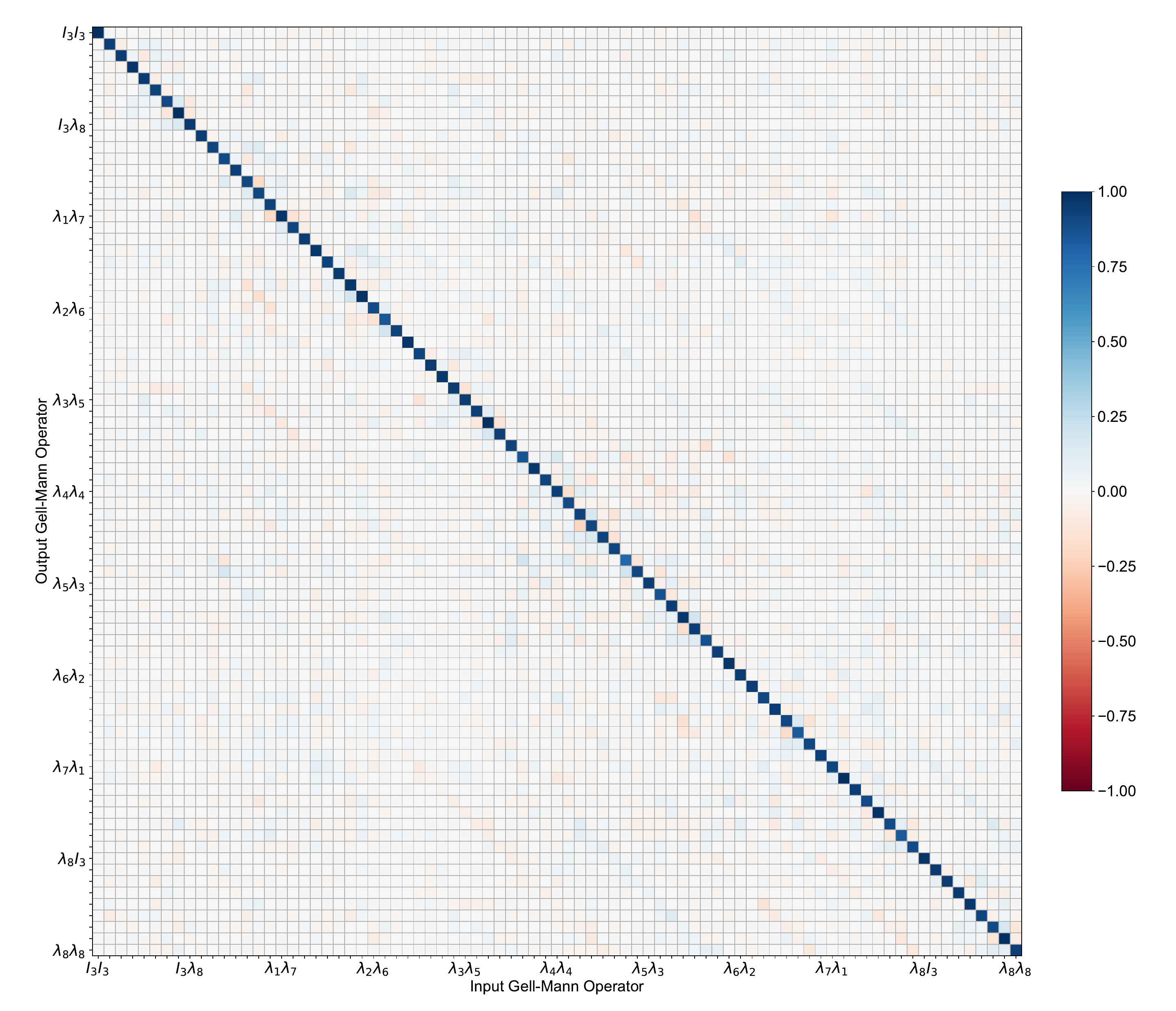}
    \caption{\textbf{Comparing the experimentally reconstructed process matrix to the ideal case}. We plot $\mathcal{E}_{\textnormal{ideal}}^\dag \mathcal{E}_{\textnormal{exp}}$, from which we can estimate the process fidelity of the CZ$^\dag$ gate.}  \label{fig:th_QPT}
\end{figure}

\noindent We analyze the Pauli transfer matrix (PTM) of the CZ$^\dag$ gate in the Gell-Mann basis.
\noindent To construct the PTM in this basis, we prepare 81 two-qutrit input states by applying an informationally complete set of native gates on each qutrit: $ \{I, \textnormal{X}^{(01)}_{\pi/2},\textnormal{Y}^{(01)}_{\pi/2},\textnormal{X}^{(01)}_{\pi},\textnormal{X}^{(12)}_{\pi}\textnormal{X}^{(01)}_{\pi}, \textnormal{Y}^{(12)}_{\pi}\textnormal{X}^{(01)}_{\pi/2}, \textnormal{X}^{(12)}_{\pi/2}\textnormal{X}^{(01)}_{\pi}, \textnormal{Y}^{(12)}_{\pi/2}\textnormal{X}^{(01)}_{\pi},\textnormal{X}^{(12)}_{\pi}\textnormal{X}^{(01)}_{\pi/2} \}$. We measure the output tomography using the same set of native gets, and reconstruct the PTM from the data using maximum likelihood estimation method. Figure \ref{fig:exp_QPT} shows the reconstructed PTM of the CZ$^\dag$ gate $\mathcal{E}_{\textnormal{exp}}$ in the Gell-Mann basis, and in Figure \ref{fig:th_QPT} we plot $\mathcal{E}_{\textnormal{ideal}}^\dag \mathcal{E}_{\textnormal{exp}}$. The process fidelity is calculated from the PTM to be $\mathcal{F}_{\textnormal{PTM}} = \textnormal{Tr}[\mathcal{E}_{\textnormal{ideal}}^\dag \mathcal{E}_{\textnormal{exp}}]/D^2 = 93.2\%$. We note that as process tomography does not decouple our characterization of our gate errors from state preparation and measurement errors, it is disfavored as a benchmarking technique when compared to CB or XEB.

\clearpage
\bibliographystyle{naturemag}
\bibliography{supplement}